\documentclass[aps,showpacs,preprint]{revtex4}
\usepackage{graphicx}
\usepackage{amssymb}

\begin{document}

\title{Renormalization group 
limit-cycles and  field theories for  elliptic S-matrices
}
\author{Andr\'e  LeClair$^{\clubsuit , \spadesuit}$ 
 and  Germ\'an Sierra$^\spadesuit$}
\affiliation{$^\clubsuit$Newman Laboratory, Cornell University, Ithaca, NY}
\affiliation{$^\spadesuit$Instituto de F\'{\i}sica Te\'orica, UAM-CSIC, Madrid, Spain} 
\date{March 2004}

\bigskip\bigskip\bigskip\bigskip

%
\font\numbers=cmss12
\font\upright=cmu10 scaled\magstep1
\def\stroke{\vrule height8pt width0.4pt depth-0.1pt}
\def\topfleck{\vrule height8pt width0.5pt depth-5.9pt}
\def\botfleck{\vrule height2pt width0.5pt depth0.1pt}
\def\Zmath{\vcenter{\hbox{\numbers\rlap{\rlap{Z}\kern
0.8pt\topfleck}\kern 2.2pt
                   \rlap Z\kern 6pt\botfleck\kern 1pt}}}
\def\Qmath{\vcenter{\hbox{\upright\rlap{\rlap{Q}\kern
                   3.8pt\stroke}\phantom{Q}}}}
\def\Nmath{\vcenter{\hbox{\upright\rlap{I}\kern 1.7pt N}}}
\def\Cmath{\vcenter{\hbox{\upright\rlap{\rlap{C}\kern
                   3.8pt\stroke}\phantom{C}}}}
\def\Rmath{\vcenter{\hbox{\upright\rlap{I}\kern 1.7pt R}}}
\def\Z{\ifmmode\Zmath\else$\Zmath$\fi}
\def\Q{\ifmmode\Qmath\else$\Qmath$\fi}
\def\N{\ifmmode\Nmath\else$\Nmath$\fi}
\def\C{\ifmmode\Cmath\else$\Cmath$\fi}
\def\R{\ifmmode\Rmath\else$\Rmath$\fi}

\begin{abstract}

The renormalization group for maximally anisotropic $su(2)$ 
current interactions in 2d is shown to be cyclic at one loop. 
The fermionized version of the model exhibits spin-charge separation
of the 4-fermion interactions and has $\Zmath_4$ symmetry.   
It is proposed that the S-matrices for these theories
are the elliptic S-matrices of Zamolodchikov and 
Mussardo-Penati.   The S-matrix parameters are related
to lagrangian parameters by matching the period of
the renormalization group.   All models exhibit two characteristic
signatures of an RG limit cycle:  periodicity of the S-matrix
as a function of energy and the existence of an infinite number
of resonance poles satisfying Russian doll scaling.

\end{abstract}


\maketitle

\vskip 0.2cm

%
%
%
%
\def\oti{{\otimes}}
\def\lb{ \left[ }
\def\rb{ \right]  }
\def\tilde{\widetilde}
\def\bar{\overline}
\def\hat{\widehat}
\def\*{\star}
\def\[{\left[}
\def\]{\right]}
\def\({\left(}		\def\BL{\Bigr(}
\def\){\right)}		\def\BR{\Bigr)}
	\def\BBL{\lb}
	\def\BBR{\rb}
%
%
\def\zb{{\bar{z} }}
\def\zbar{{\bar{z} }}
\def\frac#1#2{{#1 \over #2}}
\def\inv#1{{1 \over #1}}
\def\half{{1 \over 2}}
\def\d{\partial}
\def\der#1{{\partial \over \partial #1}}
\def\dd#1#2{{\partial #1 \over \partial #2}}
\def\vev#1{\langle #1 \rangle}
\def\ket#1{ | #1 \rangle}
\def\rvac{\hbox{$\vert 0\rangle$}}
\def\lvac{\hbox{$\langle 0 \vert $}}
\def\2pi{\hbox{$2\pi i$}}
\def\e#1{{\rm e}^{^{\textstyle #1}}}
\def\grad#1{\,\nabla\!_{{#1}}\,}
\def\dsl{\raise.15ex\hbox{/}\kern-.57em\partial}
\def\Dsl{\,\raise.15ex\hbox{/}\mkern-.13.5mu D}
%
%
\def\ga{\gamma}		\def\Ga{\Gamma}
\def\be{\beta}
\def\al{\alpha}
\def\ep{\epsilon}
\def\vep{\varepsilon}
\def\la{\lambda}	\def\La{\Lambda}
\def\de{\delta}		\def\De{\Delta}
\def\om{\omega}		\def\Om{\Omega}
\def\sig{\sigma}	\def\Sig{\Sigma}
\def\vphi{\varphi}
%
%
\def\CA{{\cal A}}	\def\CB{{\cal B}}	\def\CC{{\cal C}}
\def\CD{{\cal D}}	\def\CE{{\cal E}}	\def\CF{{\cal F}}
\def\CG{{\cal G}}	\def\CH{{\cal H}}	\def\CI{{\cal J}}
\def\CJ{{\cal J}}	\def\CK{{\cal K}}	\def\CL{{\cal L}}
\def\CM{{\cal M}}	\def\CN{{\cal N}}	\def\CO{{\cal O}}
\def\CP{{\cal P}}	\def\CQ{{\cal Q}}	\def\CR{{\cal R}}
\def\CS{{\cal S}}	\def\CT{{\cal T}}	\def\CU{{\cal U}}
\def\CV{{\cal V}}	\def\CW{{\cal W}}	\def\CX{{\cal X}}
\def\CY{{\cal Y}}	\def\CZ{{\cal Z}}

\def\rvac{\hbox{$\vert 0\rangle$}}
\def\lvac{\hbox{$\langle 0 \vert $}}
\def\comm#1#2{ \BBL\ #1\ ,\ #2 \BBR }
\def\2pi{\hbox{$2\pi i$}}
\def\e#1{{\rm e}^{^{\textstyle #1}}}
\def\grad#1{\,\nabla\!_{{#1}}\,}
\def\dsl{\raise.15ex\hbox{/}\kern-.57em\partial}
\def\Dsl{\,\raise.15ex\hbox{/}\mkern-.13.5mu D}
%
%
%
\font\numbers=cmss12
\font\upright=cmu10 scaled\magstep1
\def\stroke{\vrule height8pt width0.4pt depth-0.1pt}
\def\topfleck{\vrule height8pt width0.5pt depth-5.9pt}
\def\botfleck{\vrule height2pt width0.5pt depth0.1pt}
\def\Zmath{\vcenter{\hbox{\numbers\rlap{\rlap{Z}\kern
0.8pt\topfleck}\kern 2.2pt
                   \rlap Z\kern 6pt\botfleck\kern 1pt}}}
\def\Qmath{\vcenter{\hbox{\upright\rlap{\rlap{Q}\kern
                   3.8pt\stroke}\phantom{Q}}}}
\def\Nmath{\vcenter{\hbox{\upright\rlap{I}\kern 1.7pt N}}}
\def\Cmath{\vcenter{\hbox{\upright\rlap{\rlap{C}\kern
                   3.8pt\stroke}\phantom{C}}}}
\def\Rmath{\vcenter{\hbox{\upright\rlap{I}\kern 1.7pt R}}}
\def\Z{\ifmmode\Zmath\else$\Zmath$\fi}
\def\Q{\ifmmode\Qmath\else$\Qmath$\fi}
\def\N{\ifmmode\Nmath\else$\Nmath$\fi}
\def\C{\ifmmode\Cmath\else$\Cmath$\fi}
\def\R{\ifmmode\Rmath\else$\Rmath$\fi}

\def\barray{\begin{eqnarray}}
\def\earray{\end{eqnarray}}
\def\beq{\begin{equation}}
\def\eeq{\end{equation}}

\def\no{\noindent}

\def\gpar{g_\parallel}
\def\gperp{g_\perp}

\def\Jb{\bar{J}}
\def\dx{\frac{d^2 x}{2\pi}}

\def\rap{\beta}
\def\s{\sigma}
\def\spec{\zeta}
\def\comb{\frac{\rap\theta}{2\pi} }
\def\Ga{\Gamma}

\def\L{{\cal L}}
\def\g{{\bf g}}
\def\K{{\cal K}}
\def\I{{\cal I}}
\def\M{{\cal M}}
\def\F{{\cal F}}

\def\gpar{g_\parallel}
\def\gperp{g_\perp}
\def\Jb{\bar{J}}
\def\dx{\frac{d^2 x}{2\pi}}
\def\imag{\Im {\it m}}
\def\real{\Re {\it e}}
\def\Jbar{{\bar{J}}}
\def\kh{{\hat{k}}}

\section{Introduction}

Any  solution of the Yang-Baxter equation that is
crossing-symmetric and unitary becomes a candidate for
the factorizable S-matrix of a quantum field theory
in 2 space-time dimensions\cite{ZZ}.   Though infinite
classes of rational and trigonometric solutions related to
quantum algebras are known to 
describe  an underlying quantum field theory, there
are  no a priori principles that ensure that every S-matrix
necessarily has such an interpretation. In this regard 
the elliptic solutions provide a striking example;  though
these solutions have been known for over 25  years, it has
remained unknown whether they have a field theory description.
In this paper  a field theory will  be proposed for 
Zamolodchikov's elliptic S-matrix\cite{Zelip}, and also
for the Mussardo and Penati's scalar  S-matrix\cite{Mussardo}.

An elliptic solution to the Yang-Baxter equation was first
found by Baxter  in connection with 
the 8 vertex model of classical lattice statistical mechanics\cite{Baxter}. 
It was later shown that it could be made crossing symmetric
and unitary\cite{Zelip}.   We emphasize that the answer to the  question 
of whether there is a field theory description underlying
this S-matrix  is not contained
in the physics of the 8 vertex model, nor equivalently the 
fully anisotropic XYZ spin chain.  There is no a priori relation
between the solution of the Yang-Baxter equation that encodes
the Boltzman weights and the S-matrix of the fundamental excitations. 
Furthermore the XYZ spin chain has a non-relativistic dispersion relation,
whereas the S-matrix is relativistic.  
Certain relativistic continuum limits of the XYZ  spin chain exist
that lead to the sine-Gordon, or massive Thirring model\cite{Luther}, but the
resulting S-matrix is not elliptic but trigonometric.

 The new ingredient we use to relate the elliptic S-matrices 
to a field theory is the limit-cycle in the renormalization group (RG)
flow of the field theory\cite{BLflow,nuclear,GW,LRS,BCS,QCD,LRSc}.
   Indeed, it was
suggested by Zamolodchikov\cite{Zelip} that the field theory, 
if it exists, should be characterized by an RG limit-cycle. 
His argument was based on the periodic properties of the S-matrix
as a function of energy (at high energy) 
 and the early arguments of Wilson showing
that such periodicities are 
naturally a consequence of a cyclic RG\cite{KWilson}.
(See section II.)

An outline of this  article  is as follows.  In section II we describe
some physical signatures of an RG limit cycle in a general, model-independent
way,  extending the discussion in \cite{LRSc} to include
the role of resonance poles.    An infinite sequence of resonance
poles leading to masses with special  scaling relations
 is another characteristic
signature of an RG limit cycle, 
termed ``Russian doll'' scaling in \cite{LRS}. 
  There is a clear distinction between
massive and massless theories.  We argue that a massive theory
can only correspond to a limit cycle in the ultra-violet, whereas
a massless theory can support a limit cycle in the infra-red  {\it and} 
ultra-violet, 
in fact on all scales.   In section III  we
define the quantum field theory models as fully anisotropic
$su(2)$ current-current interactions in 2D.  The bosonized form of the
action involves both a scalar field and its dual.  The fermionized version
continues to exhibit spin-charge separation and has an explicit 
$\Zmath_4$ symmetry.     

In section IV  we show that integration of
the 1-loop beta functions leads to couplings expressed in terms
of elliptic functions, and are hence periodic as a function of scale
with the  period a function of RG invariant combinations of the couplings.  
By analyzing the $U(1)$ invariant limit we classify 4 possible models. 
  In section V 
the $\Zmath_4$ invariant 
elliptic solutions of the Yang-Baxter equation are reviewed.  
These elliptic solutions are related to  
 the field theory in sections VI and VII.
In the $U(1)$  invariant limit these theories go over
to  the usual sine-Gordon model\cite{ZZ} or
to the cyclic sine-Gordon model studied in \cite{LRS}\footnote{ 
Two different S-matrices were considered in \cite{LRS},
one periodic in rapidity, and the other with string-like
properties.  By ``cyclic sine-Gordon'' we refer to the 
first, periodic S-matrix.  The string-like S-matrix is not real analytic
and will not be relevant to the field theories of this paper.
In appendix B we describe how it compares to the other  physical
S-matrices we consider.}. 
   That the elliptic S-matrices can
have two different $U(1)$ invariant limits 
 relies on the fact that they have 
 two different trigonometric limits at complementary elliptic moduli.  
The relation between lagrangian and S-matrix parameters is
found by matching the period of the RG with the periodic S-matrix
properties.  The models also all have the expected Russian doll spectrum
of resonances.    There is another
regime that is an elliptic deformation of the sinh-Gordon
model and this is described in section VIII, where it is proposed that the
S-matrix is the scalar one considered by Mussardo-Penati\cite{Mussardo}.
Finally in section IX we consider the massless theories. 
In all cases the $U(1)$ invariant limit serves as a non-trivial check.

\section{Physical signatures of RG limit cycles}

In this section we describe in a general, model-independent fashion 
some of the signatures of an RG limit cycle, extending the 
discussion in \cite{LRSc}. 

Before proceeding, we mention  that naively,   limit cycle behavior in
a unitary theory  appears to be inconsistent with 
Zamolodchikov's c-theorem:  $c$ is a function of the running
coupling constants, so if the couplings are periodic as a function
of scale, so is $c$.   This leaves us with a puzzle since   our model is
a hermitian perturbation of a unitary conformal field theory. 
The S-matrix is also unitary in a strict sense.  
 This issue was discussed  at length
in \cite{LRSc} for a limiting case of the models here.
Though further investigation is needed regarding this issue, 
we offer the following hint as to what may be going wrong.  
Instead of Zamolodchikov's c-function which is related to 
a two-point function of stress-energy tensors,  let us consider
the c-function $c_{\rm eff}$ that determines the finite size effects;
this is the function that was studied in \cite{LRSc}. For the quantum
field on a cylinder of circumference $R$, the one-point function 
of the trace of the stress-energy tensor is a derivative of 
$c_{\rm eff}$:
\beq
\langle T_\mu^\mu \rangle_R = -\frac{\pi}{6R} \dd{c_{\rm eff}}{R}
\eeq
It trivially follows that if $T_\mu^\mu$ is positive, then
$c_{\rm eff}$ decreases with increasing $R$.  As we now argue, 
the positivity of $T_\mu^\mu$ can generally be violated in theories
with no ultra-violet fixed point and marginal perturbations, which
is precisely the situation for our models.  
  Suppose a quantum field theory
is  described by
a conformally invariant  action $S_{\rm cft}$  perturbed  by
 operators  $\CO^A$:
\beq
\label{S00}
S = S_{\rm cft}  + \sum_A  \int \frac{d^2 x}{2\pi} ~  g_A  \CO^A 
\eeq
where $g_A$ are positive couplings.  Let $\beta_A$ be the beta function for
$g_A$ and $\Gamma_A$ the scaling dimension of $\CO^A$.  Then the
beta functions (for increasing length scale)  to lowest order are 
\beq
\label{betass}
\beta_A = (2-\Gamma_A ) g_A + O(g^2) 
\eeq
The main point is that the trace of the stress-energy tensor 
recieves quantum corrections, and since it must be zero at a fixed 
point where the beta functions are zero, one must have:
\beq
\label{Tbeta}
T_\mu^\mu = \sum_A \beta_A (g)  \, \CO^A
\eeq
The above formula is well known\cite{ZamoRG} and is easily 
 verified to lowest order in conformal perturbation theory. 
Consider first a theory that can be formulated as a perturbation of
an ultra-violet fixed point  by relevant operators, which
implies $\Gamma_A < 2$.  Then the beta-functions are positive 
to lowest order.  Furthermore, for relevant perturbations, because
of the anomalous dimensions of the couplings, 
 there is often 
no higher order corrections to the beta functions since higher powers
of $g_A$ do not have the right dimension.   So in this
situation, $T_\mu^\mu$ is generally positive and the c-theorem holds.  

The above arguments clearly point to the way in which the c-theorem
can be violated. If the  $\CO^A$ are marginal, $\Gamma_A  = 2$, 
and the beta functions start at $O(g^2)$.  There are no
general constraints on the sign of the beta function.  For instance, suppose
the theory has no ultra-violet fixed point so that the couplings
increase toward the ultra-violet rather than go to zero.
  The latter implies the
beta function must be negative.  
  This in turn implies the stress-energy tensor
need not be positive and the c-theorem is violated.  For the models
below, one can check that the beta functions are not always positive.

Let us return to the properties of theories with RG limit cycles. 
Generally speaking, cyclic behavior  in the RG flow can in 
principle exist at all scales,  or can be approached  asymptotically
in the ultra-violet (UV) or infra-red (IR), the latter being UV or IR
limit-cycle behavior.  
  Once the flow
is in the cyclic regime, the couplings are periodic
\beq
\label{II.1}
g(l + \lambda) = g(l) 
\eeq
where $l= \log L$ and $L$ is the length scale.  Above,
the period $\lambda$ is fixed and model-dependent.

\def\Ecm{E_{cm}}

It is well-known that the RG leads to  scaling relations
for the correlation functions, generally referred to 
as Callan-Symanzik equations\cite{Ramond}. 
Let $G$ be an n-point  correlation function:
\beq
\label{CS.1}
G(x_1 , x_2 , .., x_n ; g, l) = \langle 
\Phi (x_1 ) \cdots \Phi (x_n ) \rangle
\eeq
The above correlation function is  computed using the renormalized
action and is thus finite and depends on the RG scale $l$.  
We have not explicitly displayed dependence on a mass parameter
 since for our models the action contains no such parameter: 
the physical mass $m$  of particles is generated dynamically 
and is a function of $g,l$.   The Callan-Symanzik equation 
expresses the independence of $G$ on the arbitrary scale $l$:
\beq
\label{CS.2} 
\(  \frac{\partial}{\partial l }  +  \beta (g) \frac{\d}{\d g} 
+ n \gamma (g) \) G(x; g, l ) = 0
\eeq
where $\gamma (g)$ is the anomalous dimension of $\Phi$.   
Let $d_\Phi$ denote the naive (engineering) dimension of $\Phi$. 
Then since $G$ has dimension $n d_\Phi$, the above equation leads
to the scaling equation:
\beq
\label{CS.3}
\( -  \frac{\partial}{\partial s }  +  \beta (g) \frac{\d}{\d g} 
+ n( \gamma (g) - d_\Phi)  \) G( e^s x; g, l ) = 0
\eeq
The above equation can be explicitly integrated:
\beq
\label{CS.4}
G( e^s x; g,l) = e^{-s n d_\Phi } ~
\exp \( {n \int_{g}^{g(s)} dg ~ \frac{\gamma (g)}{\beta (g)} } \)  
~ G(x; g(s), l ) 
\eeq
where $g(s)$ flows according to the beta-function:
$dg/ds = \beta (g) $ with $g(0)=g$.   
Letting now $s = \lambda$,  one finds
\beq
\label{CS.5}
G(e^\lambda x ; g, l ) =  e^{- \lambda n  d_\Phi } G( x; g, l ) 
\eeq
Therefore the function $G(x)/|x|^{n d_\Phi}$  is a periodic
function of $\log|x|$ with period $\lambda$.

The S-matrix is related to the Fourier transform of the 
Green's functions to momentum space, and thus also
obeys scaling relations.   
Let $S$ denote the 2-particle to 2-particle S-matrix.  For an
integrable quantum field theory in 2 space-time dimensions, 
$S$ only depends on the kinematic variable 
$\Ecm^2 = (P_1 + P_2 )^2 $ where $P_{1,2}$ are the energy-momentum
vectors of the incoming particles (for an integrable theory,
the incoming and outgoing momenta are the same).  Since the
S-matrix is a  dimensionless quantity  one obtains:
\beq
\label{II.2}
S( e^{-s} \Ecm , g  ) = S(\Ecm , g(s) )
\eeq 
If a theory is in a cyclic regime, when $s = \lambda$ the above
equation implies a periodicity in energy:
\beq
\label{II.3}
S( e^{-\lambda} \Ecm , g) = S(\Ecm , g ) 
\eeq

\subsection{Ultra-violet limit cycles}

Let us now consider a UV limit cycle.  
We specialize to 2d kinematics, and first  assume
the spectrum of particles is massive, such that the
 energy-momentum can be parameterized
in terms of a rapidity $\rap$:
\beq
\label{II.4}
E = m \cosh \rap , ~~~~~ p = m \sinh \rap 
\eeq
Above,  $m$ is the physical mass of the particles,  and as explained above, 
 is an (unknown)  function of $g,l$.    The center of mass
energy is 
\beq
\label{II.5}
\Ecm^2 = 2 m^2 ( 1 + \cosh \beta), ~~~~~~~\beta = \beta_1 - \beta_2 
~~~~~({\rm massive ~ case}) 
\eeq
The UV limit corresponds to high energies where 
$\beta$ is large and $\Ecm \approx m e^{\beta /2}$.   The relation
eq. (\ref{II.3}) then implies a periodicity in rapidity:
\beq
\label{II.6}
S(\beta - 2 \lambda ) = S(\beta )
\eeq
The above periodicity is the primary signature of a UV limit cycle
for a massive theory.

  There can exist another signature in
the UV that has 
analogies  with properties of other models
with IR  limit cycles\cite{nuclear,GW,BCS,QCD}.  Namely, if 
$\{ E_n , g, L \}$ is the spectrum of eigenvalues of 
the hamiltonian for a system of size $L$, then 
\beq
\label{Iii}
\{ E_n , g(s) , e^s L)  \} =  \{ E_n , g,  L \},
\eeq
When $s$ equals the period of one cycle $\lambda$, the above
equation shows that 
the energy spectrum at fixed $g$ should reveal 
a discrete self-similarity  as a function of $\log L$.   The manner 
in which the spectrum can  reproduce itself after one RG cycle 
is dependent on the existence of an infinite number of eigenstates 
with the ``Russian doll'' scaling behavior: 
\beq
\label{Iiii}
E_{n+1} \approx  e^\lambda E_n , ~~~~~~n=0,1,..\infty 
 ~~~~~{\rm for ~n ~ large}~~~~~~(UV)  
\eeq 
In each cycle these eigenstates reshuffle themselves such that the
$(n+1)$'th state plays the same role as the $n$'th state of the previous
cycle.  
Since we are considering a UV limit cycle, the above property
holds  at  {\it high} energies $E_n$.

For an integrable quantum field theory in infinite volume, 
 the above property can be  manifested as  
 an infinite tower of resonances
with masses that obey special scaling relations.  
  Namely, one expects
resonances of mass $M_n$, $n=0,1,..\infty$, where
\beq
\label{dolls}
M_{n+1} \approx e^\lambda ~ M_n , ~~~~~{\rm for ~ n ~ large}
\eeq
These resonances correspond to poles in the S-matrix.
The physical strip is the region $0< Im(\beta ) < \pi$, and
the only allowed poles in this region correspond to stable
bound states and  are required to be on the imaginary
axis.  In the sequel, this requirement will provide some constraints
on the models.  Poles on the  strip $-\pi < Im (\beta ) < 0$ 
with a non-zero real part correspond to unstable resonances.  
Other non-cyclic models with a {\it finite}  number of resonances 
were studied in \cite{braz,mira2,Fring}.  S-matrices with
an infinite number of resonances were considered 
in \cite{Zelip,Mussardo,Fring2}. 
Consider a pole at
$\beta = \mu - i \eta  $ with $\mu >0$ and $0<\eta < \pi$  in the S-matrix 
for the scattering of two particles of mass  $m $. 
This corresponds to a resonance
of mass $M$ and inverse lifetime $\Gamma$ where
\beq
\label{res1}
\(M- \frac{i \Gamma}{2} \)^2 = 2 m^2 ( 1 +  \cosh (\mu -i\eta )). 
\eeq
Equivalently:
\barray
\nonumber
M^2 - \frac{\Gamma^2}{4} &=& 2 m^2 ( 1 +  \cosh \mu \cos \eta),
\\ 
\label{res2}
M\Gamma &=&  2 m^2  \sinh \mu \sin\eta.
\earray
In order for $M, \Gamma$ to both be positive, both $\mu$ and
$\eta$ must be positive. 
Consider an infinite sequence of resonance poles at $\beta_n = \mu_n 
-i \eta_n $.  
When $\mu_n$ is large
one finds:
\beq
\label{largemu}
M_n \approx m e^{\mu_n / 2} ~ \cos (\eta_n /2) 
, ~~~~~
\Gamma_n \approx 2 m e^{\mu_n / 2} \sin (\eta_n /2 ) 
\eeq
Comparing with eq. (\ref{dolls}) 
one finds
\beq
\label{dolls2}
 \mu_{n+1} - \mu_n = 2 \lambda ~~~~~(UV) 
 \eeq
The above equation relating the location of the real parts of
the poles of the S-matrix is another primary signature of a UV
limit cycle.

Note that the above two signatures  of a UV limit cycle are
not independent.   If the S-matrix has resonance poles
and is also periodic in rapidity,  then the $2\lambda$ periodicity 
in rapidity automatically implies that the resonance 
poles satisfy eq. (\ref{dolls2}).   Our elliptic S-matrices
will indeed have both the periodicity in rapidity and the
Russian doll spectrum of resonances.

\subsection{Infra-red limit cycles}

Suppose the theory has a limit cycle that is  approached
in the infra-red rather than the ultra-violet.
For now let us continue
to suppose  the spectrum is massive, though as we will argue,
this is does not appear to be consistent.    Here the periodicity in 
energy eq. (\ref{II.3}), does not directly 
lead to a periodicity in $\beta$ 
since $\beta$ is small at low energies and $\Ecm \approx 2 m$.
Thus, for a massive theory,  periodicity in rapidity of the
S-matrix is {\it not} a signature of an IR limit cycle.  
  
Another possible signature is again based 
on eq. (\ref{Iii}) with $s=\lambda$, which 
leads to  a Russian doll scaling spectrum at {\it low}
energies:
\beq
\label{IRdoll}
E_{n+1} \approx e^{-\lambda} ~ E_{n}, ~~~~~~~{\rm for ~ n ~ large}
\eeq
Note the minus sign in comparison to the UV signature eq. (\ref{Iiii}). 
This corresponds to an accumulation of 
resonances  near zero energy.
Indeed the  extension of the BCS hamiltonian in \cite{BCS} and the model 
in \cite{GW} has an IR limit cycle with the property eq. (\ref{IRdoll}). 
It turns out it  is not possible to obtain a spectrum of masses 
scaling like in eq. (\ref{IRdoll})  from resonance
 poles in the S-matrix as eq. (\ref{res2}) shows:  
$\cosh \mu$ behaves always as $e^\mu$ with $\mu >0$.   

Based on the above discussion we conclude that a massive
theory cannot support a limit cycle in the IR.  This is in
accordance with the effective central charge computations in
\cite{LRSc} which show that $c_{\rm eff}$ is only quasi periodic in the UV
and  decays to zero in the 
IR  for a massive theory.    

Both the above problems are fixed if the theory is massless.   
Here the massless dispersion
relations $E = \pm p$ can be parameterized as:
\barray
E &=& \frac{m}{2} e^{\beta_R} , ~~~~~~ p = \frac{m}{2} e^{\beta_R}, 
~~~~~~~~~~~
{\rm for ~ right-movers} 
\nonumber
\\
E &=& \frac{m}{2} e^{-\beta_L} , ~~~~~ p = - \frac{m}{2} e^{-\beta_L}, ~~~~~~~
{\rm for ~ left-movers} 
\label{II.7}
\earray
where now $m$ is an energy scale.    The center of mass energy for
a right-mover with rapidity $\beta_R$ scattering with a
left-mover of rapidity $\beta_L$ is 
\beq
\label{II.8}
\Ecm = m e^{\beta /2} , ~~~~~~\beta = \beta_R - \beta_L ,
~~~~~~({\rm massless ~ case})
\eeq
If $S_{RL} (\beta)$ is the S-matrix for the scattering of 
right-movers with left-movers, then eq. (\ref{II.3}) again implies
a periodicity:
\beq
\label{II.9}
S_{RL} (\beta -  2 \lambda ) = S_{RL} (\beta ) 
\eeq

One sees then that the main difference between the massive and
massless  case is that in the massive case the periodicity in
rapidity eq. (\ref{II.6}) implies a periodicity in energy $\Ecm$ 
only for large $\Ecm$, whereas in the massless case it leads to
a periodicity at {\it all} energy scales because 
$\Ecm \propto e^{\beta/2}$.    Thus a massless theory
with the periodicity eq. (\ref{II.9}) is consistent with 
a cyclic RG flow on all scales, i.e. both the IR and UV.  

The same conclusion is reached when one considers resonances.
Consider a pole in $S_{LR} (\beta) $ at $\beta = \mu - i \eta$.   
In the massless case eq. (\ref{res1}) becomes:
\beq
\label{masslesspole}
\( M - i \frac{\Gamma}{2} \)^2 = m^2 \exp ({\mu - i \eta})
\eeq
and equation (\ref{largemu}) is exact.   The point is that unlike the
massive case, now 
$\mu$ is allowed to be negative.  An IR spectrum of masses
satisfying
\beq
\label{IRdolls3}
M_{n+1}  =  e^{-\lambda} M_n , ~~~~~n= 0, 1, ..., \infty, ~~~~~(IR)
\eeq
is possible with an infinite sequence of resonance poles satisfying:
\beq
\label{IRdolls4}
\mu_{n+1} - \mu_n = - 2\lambda, ~~~~~~~(IR)
\eeq

A massless model with resonance poles of real part $2\lambda n$ for
all $n$ positive or negative has {\it both} the IR and UV signatures
of a limit cycle and can thus describe a cyclic RG on all scales.  
The massless S-matrices we consider in the sequel have this property. 
(See section IX.)

\section{The models and their bosonic and fermionic descriptions}

\subsection{Fully anisotropic current perturbations}

We consider a conformal field theory with $su(2)$ level $\kh$  current
algebra symmetry\cite{KZ,Witten} perturbed by fully anisotropic
current interactions:
\beq
\label{2.1}
S = S_{\rm cft}  +  \int \frac{d^2 x}{2\pi} ~ 
4 \(  g_x J^x \Jbar^x + g_y J^y \Jbar^y + g_z J^z \Jbar^z \)  
\eeq
where $g_x \neq g_y \neq g_z$ are marginal couplings.  Above, $S_{\rm cft}$ is
formally the action for the conformal field theory, e.g. the
Wess-Zumino-Witten action at the critical point. 
   The currents are normalized to have
the following OPE:
\beq
\label{2.2}
J^a (z) J^b (0) \sim \frac{\kh}{2z^2}  \delta^{ab} + 
\inv{z} f^{abc} J^c (0) 
\eeq
where $z=(t+ix)/\sqrt{2}, \zbar =(t-ix)/\sqrt{2} $ 
are euclidean light-cone space-time variables. 
(Minkowski space with real time is obtained by
$t\to it$.)  
The structure constants are  $f^{abc} = i \epsilon^{abc}$, where
$\epsilon^{abc}$ is the completely antisymmetric tensor,   and similarly
for the right-moving currents $\Jbar^a (\zbar )$.

Our subsequent analysis will involve consideration of the 
$U(1)$ current with components $J^z, \Jb^z$.   We normalize
this current as follows:
\beq
\label{current}
j_\mu = (j_z , j_\zbar) = \inv{2\pi} (J , \Jb )
\eeq
and the $U(1)$ charge $T$  as
\beq
\label{u1charge}
T = \int dx  ~ j_t = \int dx ~ (j_z + j_\zbar )
\eeq
With this normalization,  the currents
$J^\pm = (J^x \pm i J^y )/\sqrt{2}$ have
$U(1)$ charge $\pm 1$.  

When the level $\kh =1$, there are simple realizations of
the current algebra in terms of a free boson or a doublet
of free fermions\cite{GoddardOlive},
 which are described in the next subsections.

\subsection{Bosonization}

\def\vphib{{\bar{\vphi}}}
\def\phid{{\tilde{\phi}}}
\def\s2{\sqrt{2}}

When $\kh=1$ the $su(2)$ currents have a free massless boson
representation with Virasoro central charge $c=1$.   The
action $S_{\rm cft}$ is just the massless Klein-Gordon 
action.  We normalize this action as follows:
\beq
\label{klein}
S_{\rm cft} = \inv{4\pi} \int d^2 x  ~~ \inv{2} \d_\mu \phi \d_\mu \phi 
\eeq
so that the propagator is:
\beq
\label{prop}
\langle \phi (z,\zbar ) \phi (0) \rangle = - \log z\zbar 
\eeq
In the conformal limit the free boson $\phi$
can be separated into its right and left moving parts, 
$\phi = \vphi (z) + \vphib (\zbar)$,  and the currents
have the following expressions:
\barray
\nonumber
J^\pm &=& \inv{\sqrt{2}} \exp ({\pm i \sqrt{2} \vphi}) , ~~~~~J^z = \frac{i}{\sqrt{2}}
\d_z \vphi 
\\
\Jb^\pm &=& \inv{\sqrt{2}} \exp ({\mp i \sqrt{2} \vphib}) , ~~~~~\Jb^z
 = - \frac{i}{\sqrt{2}}
\d_\zbar \vphib
\earray

In this bosonic representation,  the $U(1)$ current
is topological.  In Minkowski space:
\beq
\label{topcur}
j^\mu = \inv{2\s2 \pi} \epsilon^{\mu\nu} \d_\nu \phi, 
\eeq
where $\epsilon^{01} = -\epsilon^{10} = 1$,
and it  is identically conserved:  $\d_\mu j^\mu=0$.

One finally finds for the action:
\beq
\label{Sbosonic}
S = \inv{4\pi} \int ~ d^2 x ~ 
\inv{2} \( 
(1+4 g_z) (\d_\mu \phi )^2  
+ 8 g_+  \cos \s2 \phi 
+ 8 g_-  \cos \s2 \phid 
\)
\eeq
where the dual field $\phid = \vphi - \vphib$,
and $g_\pm = g_x \pm g_y$.    
Noting that $\d_z \phi = \d_\zbar \phid, 
\d_\zbar \phi = - \d_\zbar \phid$, the relation
between $\phi $ and its dual can be expressed in
Minkowski space as: 
\beq
\label{duality}
\d^\mu \phid = \epsilon^{\mu\nu} \d_\nu \phi = 2\s2 \pi ~ j^\mu
\eeq

The operators $\exp ({\pm i \s2 \phid})$ have $U(1)$ charge
$\pm 2$ and thus the $U(1)$ symmetry is broken when 
$g_x \neq g_y$.  When $g_x = g_y$ the dual field does not
appear, and the model is the sine-Gordon model with $T$ the usual
topological charge for the $U(1)$ symmetry. 
Models similar to the one defined in eq.(\ref{Sbosonic}) 
have been studied in references \cite{Boyanovsky,Aldo,Lecheminant,SierraKim},
especially the self dual cases where $g_+ = g_-$ 
and when the perturbations
of the gaussian model are relevant. 
In our model the latter perturbations 
are marginal and, as we shall see in section IV, 
the self dual constraint $g_+=g_-$ (e.g. $g_y=0$)  
is preserved by the RG only if $g_x$ or $g_z$ vanish.
Both cases correspond to well known gaussian models.

\subsection{Fermionization}

For potential applications to condensed matter physics, 
we now consider a fermionic representation of the model.  
Again when $\kh =1$, the $su(2)$ currents can be 
represented as fermion bilinears with a doublet of fermions
in the spin $1/2$ representation of $su(2)$.   
Unlike the bosonic representation above,  this is not 
an irreducible representation of the affine Lie algebra
$\hat{su(2)}$ and instead has Virasoro central charge $c=2$.  
It is thus a different model than that of the previous
subsection.   However we will show that 
  they share the same S-matrices in the interacting sector.

\def\psiL#1{\psi_{L#1}}
\def\psiR#1{\psi_{R#1}}
\def\up{\uparrow}
\def\down{\downarrow}

Introduce $su(2)$ spin-1/2  doublets of left and right moving fermions:
\beq
\label{doublets}
\Psi_L = \( \matrix{\psiL{\up}\cr \psiL{\down}\cr} \) , ~~~~~
\Psi_R = \( \matrix{\psiR{\up}\cr \psiR{\down}\cr} \)
\eeq
and their hermitian conjugates $\Psi_L^\dagger, \Psi_R^\dagger$,
where e.g. $\Psi_L^\dagger = ( \psiL{\up}^\dagger, \psiL{\down}^\dagger )$.
The currents have the following representation:
\beq
J^\pm = \inv{{2}} \Psi_L^\dagger \sigma^\pm \Psi_L , 
~~~~~J^z = \inv{2} \Psi_L^\dagger \sigma_z \Psi_L
\eeq
and similarly for $\Jb$ with $L\to R$, where
$\sigma_i$ are the standard Pauli matrices and
$\sigma^\pm = (\sigma_x \pm i \sigma_y )/2$.   
The conformal action is now:
\beq
\label{Scftferm}
S_{\rm cft}  = \int \frac{d^2 x}{2\pi} 
~ \sum_{a=\up, \down} 
\( 
\psiL{a}^\dagger \d_\zbar \psiL{a} + \psiR{a}^\dagger \d_z \psiR{a} 
\)
\eeq
The interaction terms are 
\barray
\nonumber
S_{\rm int}   &=& \int \frac{d^2 x}{2\pi} 
\Biggl[ 
  g_+  (\psiL{\up}^\dagger \psiL{\down} \psiR{\down}^\dagger \psiR{\up}
+ h.c. ) 
+  g_-  
(\psiL{\up}^\dagger \psiL{\down} \psiR{\up}^\dagger \psiR{\down} + h.c. ) 
\\
&~&~~~~~~~~~~~~~~~
+  g_z (\psiL{\up}^\dagger \psiL{\up} - \psiL{\down}^\dagger \psiL{\down} )
(\psiR{\up}^\dagger \psiR{\up} - \psiR{\down}^\dagger \psiR{\down} )
\Biggr]
\label{sintfermion}
\earray

The $U(1)$ current is now:
\beq
\label{u1fermion}
T = \inv{4\pi} \int dx 
\( 
\psiL{\up}^\dagger \psiL{\up}  - \psiL{\down}^\dagger \psiL{\down} +
\psiR{\up}^\dagger \psiR{\up}  - \psiR{\down}^\dagger \psiR{\down} \)
\eeq
With this normalization the fields have the following charges:
\barray
\nonumber 
T&=&+1: ~~~~~~ \psiL{\down} , 
\psiR{\down}, \psiL{\up}^\dagger, \psiR{\up}^\dagger
\\
\label{charges}
T&=&-1: ~~~~~~ \psiL{\up} , \psiR{\up}, 
\psiL{\down}^\dagger, \psiR{\down}^\dagger
\earray
As before, using this one sees that the $(g_x - g_y)$ terms have
charge $T=\pm 2$ and break the 
$U(1)$ symmetry.   However there
is a remaining $\Zmath_4$ symmetry.  Let $\CT_\theta$ denote
a finite $U(1)$ transformation by an angle $\theta$.  
Namely,  if an operator $\CO_q$ has $U(1)$ charge $q$, 
then $\CT_\theta (\CO_q) = e^{i q \theta} \CO_q $.   
Then, based on the charges in eq. (\ref{charges}), one
finds $\CT_\theta ( J^\pm \Jb^\pm ) = e^{\pm 4 i \theta} J^\pm \Jb^\pm $. 
Thus the action is only invariant for $\theta = \pi/2$, and
since $(\CT_{\pi/2})^4 =1$, this corresponds to a $\Zmath_4$ symmetry. 
The elliptic S-matrices in \cite{Zelip},  which we will propose
in the sequel to describe the model,  where shown there to 
be $\Zmath_4$ symmetric.

The fully anisotropic model continues
to enjoy the so-called spin-charge separation.
(For a discussion in the $U(1)$ invariant case, see
 for instance \cite{tsvelik}.)  
  The fermions can be bosonized with {\it two}
bosons $\phi_\up , \phi_\down$:
\barray
\label{fermbos}
\psiL{\up}^\dagger &=& e^{i\vphi_\up}, ~~~~\psiL{\up}= e^{-i \vphi_\up},
~~~~~\psiL{\up}^\dagger \psiL{\up} = i \d_z \phi_\up 
\\
\nonumber
\psiR{\up}^\dagger &=& e^{-i\vphib_\up} , ~~~~ \psiR{\up} = e^{i \vphib_\up}
~~~~~   
\psiR{\up}^\dagger \psiR{\up} = - i \d_\zbar \phi_\up
\earray
and the same with $\up \leftrightarrow  \down$.  Defining bosons for the
spin and charge degrees of freedom:
\beq
\label{spincharge}
\phi_s = \inv{\s2} (\phi_\up - \phi_\down ) , ~~~~~
\phi_c = \inv{\s2} (\phi_\up + \phi_\down )
\eeq
then the action is:
\beq
\label{spinch2}
S =  \inv{4\pi} \int d^2 x  ~ \[ \inv{2} (1+4 g_z) (\d\phi_s )^2 
+ \inv{2} (\d\phi_c )^2 
+ 4 g_+ \cos \s2 \phi_s  + 4 g_- \cos \s2 \tilde{\phi_s}
\] \eeq
Thus the spin and charge fields are decoupled.  The $\phi_c$ field
is a  free boson and the $\phi_s$ field has the same action as
the boson of the $c=1$ model eq. (\ref{Sbosonic}).  Thus the
S-matrices we propose below for the bosonic model
also describe the scattering of the 
spin degrees of freedom for the  fermionic model.  This explains
the $\Zmath_4$ symmetry of this S-matrix, though this symmetry
is hidden in the bosonic description.

\section{Renormalization group and classification of phases}

\subsection{Beta functions and periods}

More generally, consider 
 a conformal field theory perturbed by marginal operators 
$\CO^A$:
\beq
\label{action}
S = S_{\rm cft} +  \int \frac{d^2 x}{2\pi} ~ \sum_A  g_A \CO^A (x) 
\eeq
Assuming  the perturbing operators form
a   closed operator product expansion  in the conformal theory:
\beq
\label{2.3}
\CO^A (z,\zbar ) \CO^B (0,0) \sim \inv{z\zbar} C^{AB}_C ~ \CO^C (0,0),  
\eeq
the one-loop beta function is known to depend only on the
coefficients $C$\cite{ZamoRG}:
\beq
\label{2.4}
\beta_A \equiv \frac{dg_A}{dl} =  -\inv{2}  \sum_{B,C}  C^{BC}_A  g_B g_C 
\eeq
where $l$, the RG `time', is the log of the length scale,
and the flow toward the infra-red corresponds to increasing $l$.  
For the current-current interactions defined in eq. (\ref{2.1}),
using eq. (\ref{2.2}) one finds that the only non-zero $C$'s are 
\beq
\label{Cs}
C^{xy}_z = C^{yx}_z = C^{zx}_y = C^{xz}_y = C^{yz}_x = C^{zy}_x = -4
\eeq
This  gives
\beq
\label{2.5}
\beta_x  =  4 g_y g_z, ~~~~~
\beta_y   =  4 g_x g_z, ~~~~~
\beta_z  =  4 g_x g_y 
\eeq

\def\sn{{\rm sn}}
\def\ns{{\rm ns}}
\def\cs{{\rm cs}}
\def\ds{{\rm sn}}
\def\krg{k_{\rm rg}}
\def\krgp{k'_{\rm rg}}
\def\coth{{\rm coth}}
\def\cot{{\rm cot}}

The RG flows possess the following  RG invariants  
satisfying $\sum_i \beta_i \d_{g_i}  Q = 0$:
\beq
\label{2.6}
Q_x = g_z^2 - g_y^2 , ~~~~~
Q_y  = g_z^2 - g_x^2  , ~~~~~ Q_z  = g_x^2 - g_y^2  
\eeq
There are only two independent invariants since $Q_z = Q_x - Q_y$. 
We will thus express everything in terms of $Q_x , Q_y$. 
For the  model defined by the action (\ref{2.1}), the 
$su(2)$ symmetry is maximally broken.   
However when $g_x = g_y$ the $U(1)$ symmetry generated by
the currents $J^z , \Jbar^z$ is preserved. This symmetry 
will guide us to classify the possible models.
Observe that the self dual constraint $g_+ = g_- (g_y=0)$, 
discussed at the end of section III.B in connection with the models
in \cite{Boyanovsky,Lecheminant,SierraKim}, is not preserved
by the RG flow unless $g_x =0$ or $g_z=0$. If $g_x=0$ the model is obviously
gaussian, while if $g_z=0$ the model is also 
gaussian with exponents depending on the value of $g_x$
\cite{Lecheminant}. These models where two couplings $g's$ vanish
simultaneously correspond to degenerate situations that shall not be
consider in what follows.

Let us study the RG evolution of the coupling $g_z$
associated to this $U(1)$ symmetry.   
We can distinguish two cases:
\barray
\label{2cases}
{\rm (i)} & &  g_z^2 > g_x^2 > g_y^2~~~~~\Longrightarrow~~~~~ 
Q_x > Q_y >0 \\
\nonumber 
{\rm (ii)} & &  g_z^2 < g_x^2 < g_y^2~~~~~\Longrightarrow~~~~~
Q_x < Q_y <0 
\earray

\no The situation where
$g_x^2 > g_z^2 > g_y^2$ leads,  in the $U(1)$ limit,
to the isotropic case and it is thus contained
in the isotropic limit of (i) and (ii). 
Using the RG invariants one can eliminate $g_x , g_y$ from $\beta_z$:
\beq
\label{2.7}
\frac{d g_z}{dl} = 4 s_x s_y \sqrt{(g_z^2-Q_x)(g_z^2-Q_y) } 
\eeq
where $(s_x , s_y ) = ({\rm sign} (g_x), {\rm sign} (g_y))$.  
The solution of the above equation for  the cases (i) and (ii)
can be expressed in terms of Jacobi elliptic functions: 
\barray 
\label{2.8a}
g_z (l) & = &  \sqrt{Q_x} ~  \ns \( 4 \sqrt{Q_x} s_x s_y (l_\infty-l);  
\krg \), \qquad ~~~Q_x  > 0  \\ 
\label{2.8b}
g_z (l) & = &  \sqrt{-Q_x} ~  \cs \( 4 \sqrt{-Q_x} s_x s_y (l_\infty-l);  
\krgp \), \qquad Q_x  < 0 
\earray 
\no where $l_\infty$ is the value of $l$ at which $g_z(l_\infty) = \infty$, 
and $\ns (z ; k) \equiv 1/\sn (z;k)$, $\cs (z ; k) \equiv 
{\rm cn}(z;k)/\sn(z;k) $ are 
Jacobi elliptic functions of $z$ with modulus
$k$\cite{GR}. The modulus  of
the elliptic function  $\ns (z ; k)$  is here  denoted as  $\krg$, 
and is defined as 
\beq
\label{2.9}
 \krg^2  \equiv  { \frac{Q_y}{Q_x} }, ~~~~~ 0 \leq \krg^2 \leq   1 
\eeq
\no while the modulus of  $\cs (z ; k)$ is the complement
of $\krg$ ( $\krg'^2 = 1 - \krg^2$). The solutions (\ref{2.8a},\ref{2.8b})
can be mapped into one another using the equation
\beq
\ns(i u; k) = -i  ~ \cs(u; k'), \qquad  k^2 + k'^2 = 1
\label{ns-cs}
\eeq
\no so that (\ref{2.8a}) turns into (\ref{2.8b}) by writing
$\sqrt{Q_x} = i \sqrt{- Q_x}$. 

\def\Kb{{\bf K}}
\def\EsG{{\bf EsG}}
\def\EcsG{{\bf EcsG}}

The functions $\ns (z ; k), \cs (z ; k) $ have double periodicity
\barray
\label{2.10}
\ns (z + 4 m \Kb    + 2 i n \Kb'    ; k ) & = &  \ns (z ; k) \\ 
\nonumber 
\cs (z + 2  m \Kb    + 4 i n \Kb'    ; k ) & = &  \cs (z ; k)
\earray
where $m,n$ are integers, $\Kb (k)$ is the complete
elliptic integral, and $\Kb' (k) = \Kb (k') $, 
with $k' = \sqrt{1-k^2 }$.  The coupling $g_z$  is thus a periodic
function of $l$, with a  period depending  on the sign of $Q_x$:

\bigskip\bigskip

\no (i) {$\bf Q_x >0$.}    ~~~Here $\sqrt{Q_x}$ is real and 
\beq
\label{2.11}
g_z (l + \lambda_1 ) = - g_z (l) , ~~~~~ \lambda_1
 \equiv \frac{  \Kb (\krg )}
{2 \sqrt{Q_x}}
\eeq

\bigskip\bigskip

\no (ii)  {$\bf Q_x<0$.}   ~~~  Here $\sqrt{Q_x}$ is imaginary and
\beq
\label{2.12}
g_z (l + \lambda'_1) = g_z (l) , ~~~~~\lambda'_1 \equiv 
\frac{\Kb' (\krg )}{2 \sqrt{-Q_x}}
\eeq
The 1-subscripts on
$\lambda, \lambda'$ refer to being the 1-loop result. 
The reason for the extra minus sign in eq. (\ref{2.11}) in 
comparison to eq. (\ref{2.12}) will be explained below. 

\subsection{$U(1)$ invariant limits}

In the $U(1)$ invariant  limit  $g_x^2  = g_y^2 $, 
 we have $Q_x = Q_y \equiv Q$
and $\krg =1$.  The solutions of the one-loop renormalization
group again depend on the sign of $Q$:

\bigskip\bigskip

\no (i) ${\bf Q>0, \krg = 1}$

\beq
\label{2.13}
g_z (l) = \sqrt{Q}~ \coth \( 4 \sqrt{Q}~ (l_\infty - l ) \)
\eeq
The model is well understood in this limit. 
Define $\sqrt{Q}$ to be positive. 
There are two subcases depending on the sign
of $g_z$.   {\bf Massive case:} 
When $g_z >0 $, $g_z$ flows to an ultra-violet fixed
point at $g_z = \sqrt{Q}$ as $l \to - \infty$ and to a strong
coupling fixed point in the infrared;  
it is thus 
a massive theory. {\bf Massless case:}  When $g_z < 0$, the fixed point is
at $g_z = -\sqrt{Q}$ in the infrared ($l \to + \infty$).  
This is thus a massless theory. Both these theories 
have a sine-Gordon description where the conventional sine-Gordon
coupling is related to $Q$;  non-perturbative formulas were 
obtained in \cite{BLflow}, and will be used below.
   In this limit, since $\Kb(k) \approx 
\log 4/\sqrt{1-k^2} $ as $k\to 1$,  the period $\lambda_1$ 
in eq. (\ref{2.11}) goes to $\infty$, as expected for theories
with fixed points.

\no (ii) ${\bf Q < 0, \krg = 1}$ ~~~Here since $\sqrt{Q}$ is
imaginary, as in \cite{LRS} let us define:
\beq
\label{2.14}
\sqrt{Q} \equiv i h / 4 
\eeq
The one-loop RG flow is now
\beq
\label{2.15}
g_z (l) = \frac{h}{4} ~ \cot ( h (l_\infty-l)) 
\eeq
In this case the periodicity eq. (\ref{2.12}) is maintained. 
Since $\Kb' (k) \approx \pi/2$ as $k\to 1$, the period $\lambda'_1$
becomes
\beq
\label{2.16}
\lambda'_1 \to  \frac{\pi}{h}       ~~~~~               (\krg = 1)
\eeq
This agrees with the manifest periodicity in eq. (\ref{2.15})
and the one-loop period computed in \cite{LRS}.    

In the $U(1)$ invariant limit the higher order contributions
to the beta functions are known\cite{GLM}.   These beta functions
were used to classify the various phases as a function of $\sqrt{Q}$
in \cite{BLflow}.  The various regimes are distinguished by whether
the fixed points are in the IR or UV, or whether the flow is cyclic. 
  The result is summarized in figure 1.  

\begin{figure}[h]
\begin{center}
\includegraphics[height=6 cm, angle=0]{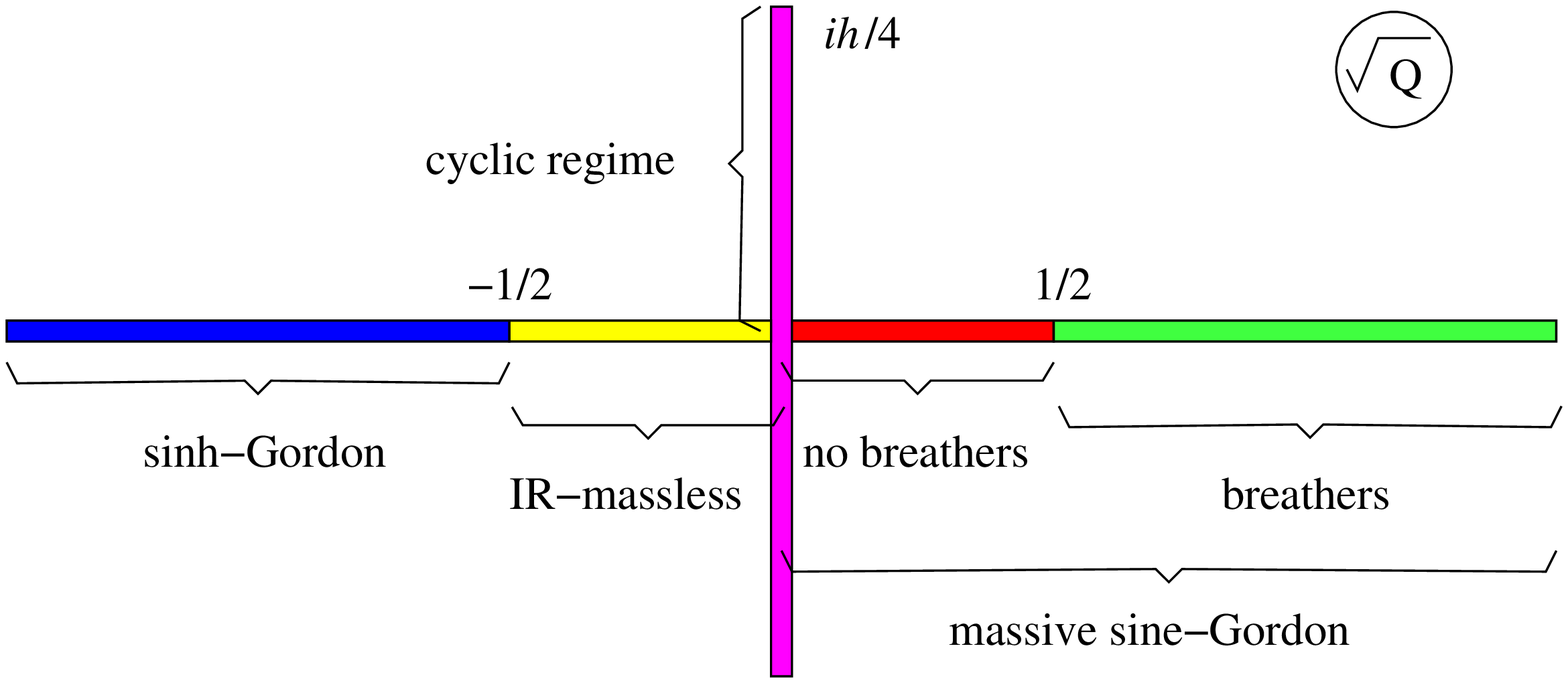}
\end{center}
\caption{Different regimes in the $U(1)$ invariant limit 
 as a function of $\sqrt{Q}$. } 
\label{fig1}
\end{figure}

In order to classify the possible models when $\krg \neq 1$,  we
assume that each distinct regime of the $U(1)$ invariant model 
has an elliptic deformation that is integrable. 
This is one of the main hypotheses of this paper. 
  As we will show in the sequel, the
fact that consistent  S-matrices can be proposed for all these regimes
supports this hypothesis.   Our nomenclature for the models
refers to the $\krg =1$ limit, and  $Q$ refers to  $Q_x$ in this
limit.     We can thus identify the following
distinct models:

\def\EshG{{\bf EshG}}
\def\mEsG{{\bf mEsG}}

\bigskip

\no 
{\bf Elliptic sine-Gordon model} ($\EsG$).  This is an elliptic
deformation of the usual sine-Gordon model with  
$0 \leq \sqrt{Q} < \infty  $.  

\bigskip

\no {\bf Elliptic cyclic sine-Gordon model} ($\EcsG$).  This is
an elliptic deformation of the cyclic sine-Gordon model described
in \cite{LRS} where $Q <0   $ and $\sqrt{Q} = i h/4 $,  with $h>0$.  

\bigskip

\no {\bf Massless elliptic sine-Gordon model}  ($\mEsG$ ). 
Here $-1/2 \leq \sqrt{Q} \leq 0$.    This massless model
is characterized by having an infrared fixed point in the $U(1)$ 
invariant limit.

\bigskip 

\no {\bf Elliptic sinh-Gordon model}  ($\EshG$).   Here
$-\infty < \sqrt{Q} < -1/2$ and the model reduces to the usual sinh-Gordon
model with one massive scalar particle in the $U(1)$ invariant
limit.

\bigskip

\def\lamsG{\lambda}
\def\lamcsG{\lambda'}

It is important to note 
that under permutations of the labels $x,y,z$, the models 
can be mapped into each other, and this provides certain
consistency checks of our results. Consider first the 
permutation:
\beq
\label{perm0}
g_x \leftrightarrow g_y \Longrightarrow 
Q_x \to \krg^2 Q_x ,  ~~~~~\krg \to 1/\krg 
\eeq
Under this transformation, the model is mapped into itself,
which means that the period of the RG $\lambda_1 , \lambda'_1$
should be invariant.   This is readily checked using
$Re(\Kb (1/k) ) = k \Kb (k)$.  
Consider next the permutation:
\beq
\label{perm1}
g_y \leftrightarrow g_z \Longrightarrow ~~~Q_x \to - Q_x , ~~~ \krg \to \krg'
\eeq
This implies that the models $\EsG$ and $\EcsG$ are essentially
the same at complementary elliptic moduli.   
In particular, 
   when $g_x^2 = g_z^2$,  $Q_y =0$,
and thus $\krg =0$ is another $U(1)$ invariant limit of $\EsG$. 
Since the sign of $Q_x$ is flipped, 
this means  that the $\EsG$ model at
$\krg = 0$ can be mapped onto the  
 $\EcsG$ at $\krg =1$, and visa versa. 
The extra minus sign in eq. (\ref{2.11}) was chosen so that
the periods $\lamsG , \lamcsG$ agree under this exchange.  
 In the sequel this will
serve as an important consistency check of the S-matrices. 
Though the  two
models $\EsG$ and $\EcsG$ are essentially the same,  
S-matrix descriptions given below are different 
because as defined the models have different $U(1)$ limits
as $\krg =1$, one being the usual sine-Gordon model, the
other the cyclic sine-Gordon model studied in \cite{LRS}. 
However since the $U(1)$ limit at $\krg = 0$ does not 
correspond to the conventional sine-Gordon action, but only
after the permutation $\CP ( g_z ) = g_y, \CP (g_y) = g_z$, 
as we will see the S-matrices in this limit match onto known
$U(1)$ invariant S-matrices up to a transformation $\CP$ with
$\CP^2 = 1$.  
Finally consider:   
\beq
\label{perm2}
g_x \leftrightarrow g_z \Longrightarrow ~~~ 
Q_x \to \krg'^2 Q_x , ~~~~~\krg \to i \frac{\krg}{\krg'}
\eeq
Here since $\krg$ becomes imaginary,  this does not lead to any
equivalence between models.

The above analysis is only at one loop and one must investigate
whether the main features persist to higher orders.
Our basic assumptions in the sequel are the following.     
First,  the one-loop RG invariants $Q$,
when appropriately corrected to higher orders, must
continue to be RG invariants.  Secondly,  the RG flows must
continue to be cyclic where the period of the RG, $\lambda (Q)$,  is
a function of the  higher order corrected $Q$'s.    In the 
$U(1)$ invariant limit, this is precisely the situation\cite{BLflow, LRS}.  
When $g_x^2 = g_y^2$ one has: 
\beq
\label{higher1}
Q_x = Q_y = \frac{g_z^2 - g_x^2}{(1- g_z)^2 (1-g_x^2 ) } ~~~~~~(g_y^2= g_x^2)
\eeq
The higher order 
 beta-functions can also be integrated exactly in the $U(1)$ limit,
and the period of
the RG in the cyclic regime is $\lambda = \pi/2 \sqrt{-Q_x}$,  which
is precisely twice the 1-loop result\cite{LRS}.

In Appendix A, we study the higher order corrections and provide
analytical and numerical evidence for the above hypotheses.  
There we give evidence that in the fully anisotropic case the
period is again twice the one loop result.
This leads us to define the RG periods 
$\lamsG, \lamcsG$ for the models $\EsG, \EcsG$ respectively:
\beq
\label{lams}
\lamsG \equiv 2 \lambda_1 , ~~~~~ \lamcsG \equiv 2 \lambda'_1
\eeq
where $\lambda_1 , \lambda'_1$ as functions of $Q_{x,y}$ are given 
in eqns. (\ref{2.11},\ref{2.12}) and the $Q$'s are higher loop
corrected ones.    The fact that we can find exact S-matrices
with the expected properties is a good  indication of the validity
of our hypotheses.

In the above 1-loop RG analysis,  the couplings are periodic
on all length scales.  The arguments given in section II suggest
that all the elliptic models should be massless.  This would however
be inconsistent with some of the above $U(1)$ limits which are known
to be massive, unless a mass develops dynamically in going to this limit. 
It may also be that depending on what regularization one in practice 
uses to define the  models, for example a lattice cut-off,  the limit
cycle may be only observable in the UV or IR.  For this reason in
the sequel we consider both the massive and massless cases.  
Most of the discussion will focus on the massive case
since these results are straightforwardly extended to the massless
case since $S_{LR}$ is the same function as in the massive case.  
(See section IX.)

\def\S{\Sigma}

\section{Non-diagonal Elliptic S-matrices}

In this section we review Zamolodchikov's 
elliptic S-matrix\cite{Zelip},
defering its relation to the quantum field theory to the next section.   
As usual,  introduce relativistic massive dispersion 
relations eq. (\ref{II.4}). 
The S-matrices are more clearly presented in a real basis for the
particles. 
Let $A_1 , A_2$ formally denote creation operators for the two
particles in the theory.  The S-matrix is encoded in the exchange
relation:
\beq
\label{3.2}
A_a (\beta_1 ) A_b (\beta_2 ) = \Sigma_{ab}^{cd} (\beta_1 - \beta_2 ) 
A_d (\beta_2 ) A_c (\beta_1 ) 
\eeq
The S-matrix $\S$ does not possess a $U(1)$ symmetry and thus has some
additional non-zero amplitudes in comparison to e.g. the sine-Gordon
S-matrix.    Define:
\barray
\nonumber
\sigma &=& \S_{11}^{11} = \S_{22}^{22}
\\
\nonumber
\sigma_t &=& \S_{12}^{12} = \S_{21}^{21}
\\
\label{3.3}
\sigma_r &=& \S_{12}^{21} = \S_{21}^{12} 
\\
\nonumber
\sigma_a &=& \S_{11}^{22} = \S_{22}^{11} 
\earray
Crossing symmetry in this basis reads:
\beq
\label{3.4}
\sigma (\beta ) = \sigma(i \pi - \beta), ~~~
 \sigma_t  (\beta ) = \sigma_t (i \pi - \beta), ~~~
\sigma_r  (\beta ) = \sigma_a (i \pi - \beta)
\eeq

The general solution to the constraints of the Yang-Baxter equation
and crossing symmetry has two free parameters, 
$k , \alpha$: 
\barray
\nonumber
\sigma_r (\beta) &=& \frac{ \sn (2\pi i \alpha - 2\alpha \beta; k)}
{\sn ( 2\pi i \alpha;k )}  ~ \sigma (\beta)
\\
\label{3.5}
\sigma_a (\beta) &=& \frac{ \sn (2\alpha \beta ;k) }{\sn (2\pi i \alpha; k )} 
~\sigma (\beta) 
\\
\nonumber
\sigma_t (\beta ) &=& 
- k ~\sn ( 2\alpha \beta ;k  )~ \sn ( 2\pi i \alpha - 2\alpha\beta ;k )
 ~ \sigma(\beta)
\earray 
(The parameter $\eta$ in \cite{Zelip} 
is here expressed as $\eta = -i \pi \alpha$.)

When the S-matrix is real analytic, 
$(\S(\beta))^\dagger = \S(-\beta )$,
 unitarity reads 
$\S (\beta ) \S (-\beta ) = 1$.  This gives the additional constraint
on $\sigma$:
\beq
\label{3.6}
\sigma (\beta ) \sigma (-\beta ) = \frac{ \sn^2 (2 \pi i \alpha ;k  )}
{\sn^2 (2\pi i \alpha ;k ) - \sn^2 (2 \alpha \beta ;k ) }
\eeq
The so-called minimal solution to the above equation and crossing symmetry
is
\beq
\label{3.7}
\log \sigma = 4 \sum_{n=1}^\infty \inv{n} 
\frac{ \sinh^2 \( {2\pi n (\pi - \gamma)}/{\gamma'} \) 
\sin \( {2\pi n \beta}/{\gamma'}\) 
\sin \( { 2\pi n (i\pi - \beta )}/{\gamma'} \) }
{\sinh \({4\pi n \gamma}/{\gamma'} \) \cosh \( {2 \pi^2 n}/{\gamma'} \)
}
\eeq
where 
\beq
\label{3.8}
\gamma \equiv \frac{\Kb' (k)}{2\alpha} , ~~~~~~
\gamma'  \equiv \frac{2 \Kb (k)}\alpha 
\eeq

The above S-matrix is real analytic so long as $\alpha$ is real. 
  However  
the infinite sum in eq. (\ref{3.7}) is convergent only  
if $\gamma >\pi/2$ when $\gamma' $ finite.  
For $0<k<1$, convergence of $\sigma$ thus requires
that $\alpha$ be real and  positive.

\def\ell{{\tilde{k}}}
\def\alphap{{\tilde{\alpha}}}
\def\xp{{\tilde{x}}}
\def\yp{{\tilde{y}}}

In order to study $U(1)$ invariant limits of the elliptic S-matrix,
we go to a complex basis of particles:
\beq
\label{S.1}
{A}_\pm = \inv{\sqrt{2}} \( A_1 \pm i A_2 \)
\eeq
The S-matrix  for the ${A_\pm }$ particles is defined as 
\beq
\label{S.2} 
A_a (\beta_1 ) A_b (\beta_2 ) = {S}_{ab}^{cd} (\beta_1 - \beta_2; k,\alpha )
~ A_d (\beta_2 ) A_c (\beta_1 ) 
\eeq
The non-zero amplitudes are 
\barray
\nonumber
S_0 &\equiv& S_{++}^{++} = S_{--}^{--}
\\
\nonumber 
S_t &\equiv& S_{+-}^{+-} = S_{-+}^{-+}
\\
\label{S.3}
S_r &\equiv& S_{+-}^{-+} = S_{-+}^{+-}
\\
\nonumber 
S_a &\equiv& S _{++}^{--} = S_{--}^{++}
\earray
The relation between $S$ and $\Sigma$ follows from eq. (\ref{S.1}):
\barray
\nonumber
2 S_0 &=& \sigma - \sigma_a + \sigma_t + \sigma_r 
\\
\nonumber
2 S_t  &=& \sigma + \sigma_a + \sigma_t - \sigma_r
\\
\label{S.4}
2 S_r &=& \sigma + \sigma_a - \sigma_t + \sigma_r
\\
\nonumber
2 S_a &=& \sigma - \sigma_a - \sigma_t - \sigma_r
\earray
Equivalently:
\barray
\nonumber
2 \sigma & = & S_0 + S_t + S_r + S_a  \\
\label{S.4bis}
2 \sigma_t  & = & S_0 + S_t - S_r - S_a  \\
\nonumber
2 \sigma_r & = & S_0 - S_t + S_r - S_a  \\
\nonumber
2 \sigma_a & = & - S_0 + S_t + S_r - S_a  
\earray
Using (\ref{3.5}) in  eq.(\ref{S.4})  we obtain 
\barray
\nonumber 
S_t (\rap)& = & \frac{\sn( 2  \alphap  \beta  ;\ell)}{ 
\sn( 2 \pi i \alphap - 2\alphap \beta ; \ell) } ~ S_0(\rap)   \\
\label{Sexplicit}
S_r(\rap)& = & \frac{\sn(2 \pi i \alphap ;\ell)}{ 
\sn( 2 \pi i \alphap - 2 \alphap \beta  ; \ell)}~ S_0(\rap) \\
\nonumber 
S_a(\rap)& = & \ell ~ \sn( 2 \alphap\rap; \ell)~ 
\sn( 2 \pi i \alphap ;\ell)~ S_0(\rap)
\earray
\no where the modulus $\ell$ and $\alphap$ are given in terms
of the moduli $k$ and $\alpha$ by, 
\beq
\ell = \left( \frac{ 1 - \sqrt{k}}{1 + \sqrt{k}} \right)^2, 
\qquad \alphap = \frac{i}{2} ( 1 + \sqrt{k})^2 ~ \alpha
\label{ell}
\eeq
For future reference, we note the identities:
\beq
\label{gammaid}
(1+\sqrt{\ell})^2 \Kb (\ell) = \Kb' (k), ~~~~~
(1+\sqrt{\ell})^2 \Kb' (\ell) = 4 \Kb (k) 
\eeq

The proof of eq.(\ref{Sexplicit}) uses  standard tools in the theory
of doubly periodic meromorphic functions \cite{Baxter}. 
As an example let us consider
the relation 
\beq
\frac{\sigma_r}{\sigma} = \frac{ 
 S_0 - S_t + S_r - S_a}{ S_0 + S_t + S_r + S_a}
\label{proof1}
\eeq
which follows from eq.(\ref{S.4bis}). Using eqs.(\ref{3.5}) and 
(\ref{S.4}) this equation becomes,
\beq
\frac{ \sn(x-y;k)}{\sn(x;k)} =
\frac{ \sn(\xp;\ell) - \sn(\yp;\ell) + \sn(\xp - \yp;\ell) 
- \ell  \sn(\xp;\ell) \sn(\yp;\ell)  \sn(\xp - \yp;\ell)}{
 \sn(\xp;\ell) + \sn(\yp;\ell) + \sn(\xp - \yp;\ell) 
+ \ell  \sn(\xp;\ell) \sn(\yp;\ell)  \sn(\xp - \yp;\ell)}
\label{proof2}
\eeq
where 
\beq
x = 2 \pi i \alpha, ~ y = 2 \alpha \beta, ~ 
\xp =  2 \pi i \alphap, ~ y = 2 \alphap \beta
\label{proof3}
\eeq
One can easily check that the LHS and the RHS of (\ref{proof2}),
viewed as functions of $x$ or $y$, 
have the same periodicity properties, position of poles 
and zeros, and hence  by  Liouville's theorem should be 
proportional up to a constant, whose value is actually one.

In this basis,  $A_+$ and $A_-$ are charge conjugates
and crossing symmetry reads:
\beq
\label{S.5}
S_t (\beta ) = S_0 (i\pi - \beta ), ~~~
S_r (\beta ) = S_r (i\pi - \beta ), ~~~
S_a (\beta ) = S_a (i\pi - \beta )
\eeq

\def\Ah{\hat{A}}
\def\Sh{\hat{S}}

It turns out that to describe all the field theories in 
section II,  it is convenient to introduce a different
description of the  S-matrix.  Define new
particles $\Ah^\pm$ with the S-matrix exchange relation:
\beq
\label{Shat}
\Ah_a (\beta_1 ) \Ah_b (\beta_2 ) = 
{\Sh}_{ab}^{cd} (\beta_1 - \beta_2; k,\alpha )
~ \Ah_d (\beta_2 ) \Ah_c (\beta_1 ) 
\eeq
Define the non-zero amplitudes $\Sh_{0,t,r,a}$ as 
in eq. (\ref{S.3}), e.g. $\Sh_0 \equiv \Sh_{++}^{++} = \Sh_{--}^{--}$, etc,
and let:
\beq
\label{Shat2}
\Sh_0 = \sigma_r , ~~~
 \Sh_t = \sigma_a , ~~~\Sh_a = \sigma_t , ~~~ \Sh_r = \sigma
\eeq
with the $\sigma$'s the same as in eq. (\ref{3.5}).   
Then it follows from the crossing symmetry relations on the $\sigma$'s,  
eq. (\ref{3.4}),  that the $\Sh$'s satisfy the crossing relations
eq. (\ref{S.5}) with $S\to \Sh$.   Thus the S-matrix $\Sh$ 
is a proper S-matrix with $\Ah_+ , \Ah_-$ charge conjugates.

Expressing $\Sigma$ as the matrix:
\beq
\label{Sigmamat}
\Sigma = \( \matrix{\sigma &0&0&\sigma_a \cr
                    0&\sigma_t &\sigma_r & 0 \cr
                    0&\sigma_r &\sigma_t &0\cr
                    \sigma_a&0&0&\sigma \cr } \)
\eeq
then in the sequel we will express $\Sh$ as the following transformation 
of $\Sigma$:
\beq
\label{flip}
\Sh = \( 
\matrix{
\Sh_0 &0&0&\Sh_a\cr
0 &\Sh_t &\Sh_r &0 \cr
0 & \Sh_r &\Sh_t &0 \cr
\Sh_a &0&0&\Sh_0 \cr 
} \)
= \CP( \Sigma ) =  
\( \matrix{\sigma_r  &0&0&\sigma_t \cr
                    0&\sigma_a &\sigma & 0 \cr
                    0&\sigma  &\sigma_a &0\cr
                    \sigma_t&0&0&\sigma_r \cr } \)
\eeq
where $\CP^2 =1$. 
The $\CP$ transformation can be written in matrix form
as follows:
\beq
\label{flip2}
\Sh = P_2 \,  \Sigma \,  P_1
\eeq
with 
\beq
\label{flip3}
P_1  = \sigma_x \otimes 1 = \( 
\matrix{ 0&0&1&0\cr 0&0&0&1\cr 1&0&0&0\cr 0&1&0&0 \cr } \) ,
~~~~~
P_2  = 1 \otimes \sigma_x  = \( 
\matrix{ 0&1&0&0\cr 1&0&0&0\cr 0&0&0&1\cr 0&0&1&0 \cr } \)
\eeq
where $\sigma_x = \( \matrix{0&1\cr 1&0\cr} \) $ interchanges
the two particles.  Incidently, it is clear from eq. (\ref{flip2})
that if $\Sigma$ satisfies the Yang-Baxter equation, then so
does $\Sh$.    It should be kept in mind
that the $\CP$ transformation is not a change of basis.

\section{Matching lagrangian and S-matrix parameters for $\EsG$}

As shown in section II, the  field  theories are $\Zmath_4$ symmetric
and this should be a symmetry of the S-matrix.  As shown in
\cite{Zelip}, this is indeed a symmetry of the elliptic S-matrices of
the last section.   The S-matrix parameters $k, \alpha$ are dimensionless
parameters and thus must be RG invariant functions 
of the coupling constants $g_{x,y,z}$, equivalently functions
of $\krg$ and $Q_x$.  
By matching the periodicity of the RG with
the periodicity of the S-matrix we now relate the S-matrix
parameters $k,\alpha$ of the last section to the lagrangian parameters.
In this section we do this for the $\EsG$ model.  We assume the model
to be massive;  the massless version will be described in section IX.

\subsection{The S-matrix}

The S-matrices  in eqs. (\ref{3.2},\ref{3.3},\ref{Shat2}) 
enjoy the following periodicity in rapidy:
\beq
\label{3.9}
S(\beta - \gamma'  ) = S(\beta ), ~~~~~\Sh (\beta - \gamma' ) = \Sh (\beta)  
\eeq
where $\gamma'$ is defined in eq. (\ref{3.8}). Based on the results
of section III,   we match this periodicity in rapidity with 
the periodicity of the RG and  
thus  identify 
\beq
\label{3.10}
\gamma'  = 2 \lamsG
\eeq
where $\lamsG$ is the period of the RG defined in section II.  
The above equation implies
\beq
\label{3.11}
\frac{\Kb (k)}{\alpha} = \frac{   \Kb (\krg )}{\sqrt{Q_x}} 
\eeq

Eq. (\ref{3.11}) is a single equation for two parameters, however
we can argue as follows to fix them both.  
When $\krg $ is 0 or 1, then the S-matrix must be trigonometric,
i.e. $k$ must also be 0 or 1.  However when  $\krg =1$,   $k$ must
also be $1$,  otherwise the relation between $\alpha$ and $Q_x$ 
would be infinite.   This also implies that $\krg = 0$ corresponds
to $k=0$.  This suggests that $k$ is only a function of $\krg$, 
since both are between $0$ and $1$.  
 A further constraint is provided by requiring {\it both} 
 $U(1)$ invariant limits to be correct.
Comparing the $U(1)$ limits of the elliptic S-matrix to the usual
sine-Gordon and cyclic sine-Gordon ones,  this requires that 
when $k=1$, $\alpha =\sqrt{Q}/2$ and when $k=0$,  $\alpha = \sqrt{Q}$.      
We have found the following solution to all these constraints:
\beq
\label{3.12}
\sqrt{Q_x} = (1+k) \alpha, ~~~~~ \krg = \frac{2 \sqrt{k}}{1+k}   ~~~~~(\EsG) 
\eeq
We have used the identity:
\beq
\label{ident2}
\Kb \( \frac{ 2 \sqrt{k}}{1+k}  \) = (1+k) \Kb (k)
\eeq 
In summary we propose the S-matrix for $\EsG$ is:
\beq
\label{summar1}
S_{\EsG}  (\beta ) = S(\beta; k,\alpha),
\eeq
where $S(\beta; k , \alpha)$ is given in eqs. (\ref{S.3},\ref{S.4}),
and $k,\alpha$ are related to the lagrangian parameters 
$Q_x , \krg$ by eq. (\ref{3.12}).

\subsection{Resonance poles}

Above, we have related the S-matrix and lagrangian parameters
by matching the periodicity  properties of the S-matrix with the RG. 
In this subsection we show that the model also has the
resonance poles with Russian doll scaling anticipated in section II.

The function $\sn (z ; k)$ has the following zeros and poles:
\barray
\sn (z ; k) :   ~~~~~~~~~~~~~~&~&{\rm zeros:}  ~~z= 2n\Kb + 2 i m \Kb' 
\nonumber \\ 
\label{zeropoles}
&~& {\rm poles: } ~~ z = 2n\Kb + (2m+1) i \Kb' 
\earray
where $m,n\in \Zmath$. 
The zeros of $\sn (2 \pi i \alphap  - 2\alphap \beta ; \ell)$ thus  lead to 
poles of $S_{t,r}$ at:
\beq
\label{shpoles}
\beta = i \pi (1-m\gamma / \pi ) + n \gamma' 
\eeq
where we have used the identity eq. (\ref{gammaid}).  

Requiring that there are no complex poles on the physical strip
$0 < Im (\beta) < \pi$ leads to the constraint:
\beq
\label{const}
 \gamma >   \pi
\eeq
   Note
that the latter is compatible with the convergence of $\sigma$, 
which requires $\gamma > \pi / 2$.  
When $\gamma > \pi$,  there remain  resonance poles on the 
 strip $-\pi < Im (\beta ) < 0$:
\beq
\label{respoles}
\beta_n = \mu_n - i \eta_n , ~~~~~\mu_n = n \gamma' , ~~~\eta_n = \gamma - \pi
\eeq
Here, since the model is assumed massive, as explained in 
section II, $n$ must be positive.  
When $\gamma > 2 \pi$ there are no resonances since $\eta_n > 1$.  
Finally,  since  $\gamma' = 2 \lamsG$, one sees that the spectrum 
of resonances precisely satisfies the Russian doll scaling
property eq.  (\ref{dolls2}). 
That the S-matrix has both of the  UV signatures of a cyclic
RG described in section II,
 with compatible period, relies on the special relation between
the periods and poles enjoyed by Jacobi elliptic functions.

\subsection{$U(1)$ invariant limits}

Checks  of the above S-matrix  are the  $U(1)$ invariant
limits $\krg =0,1$.
This check is non-trivial because the lagrangian and S-matrix
parameters were related by only matching the periodicity
of the S-matrix, and in the $\krg =1$ sine-Gordon limit
this periodicity disappears.  The limit $\krg =0$ on the other
hand preserves the cyclicity of the  RG.  As we now show the S-matrix
gives the expected limit in both cases despite the fact that
these two limits have very different properties.

   We first begin with $\krg =1$, 
 which is expected to be the sine-Gordon model. 
The latter is defined by the lagrangian
\beq
\label{sgL}
S = \inv{4\pi} \int  d^2 x \( \inv{2} (\d\phi)^2 + \Lambda
\cos ( b \phi ) \) 
\eeq
with  $0<b^2 < 2$. 
In the  limit $\krg =1$,  $Q_x = Q_y \equiv Q$, and $k=1$.
  The relation
between $b$ and $Q$ can be found by matching the slope
of the beta function at the fixed point and is known\cite{BLflow}:
\beq
\label{bQ}
2 \sqrt{Q} = \frac{2-b^2}{b^2}
\eeq
Note that in this limit the sine-Gordon coupling $g$ depends 
on both $g_z $ and $g_x$, not only on $g_z$ as one would naively
think.  The reason is that both $g_z$ and $g_x$ continue to
flow under RG. On the other hand $b$, being a dimensionless
parameter in the S-matrix, must be an RG invariant, as  in 
eq. (\ref{bQ}). This was explained in more detail in \cite{BLflow}. 
In this limit,  $\gamma' \to \infty$,  and thus the RG period
$\lamsG$ becomes infinite, consistent with a theory
with a fixed point.  The resonances become infinitely
heavy and decouple.  

The parameter $\gamma$ remains finite and depends on the
sine-Gordon coupling constant $b$:
\beq
\label{sGparam}
\gamma = \frac{\pi}{4 \alpha} = \frac{\pi}{2 \sqrt{Q}} = 
\frac{\pi b^2}{2-b^2} 
\eeq
Since $\gamma' = \infty$,  the constraint $\gamma > \pi$ 
is not required and the whole range of the massive sine-Gordon
model $0<b^2 < 2$ is covered. 

Since $\ell = 0$ in the limit, using 
 $\sn (z; 0) = \sin z$, one finds from (\ref{Sexplicit}) 
$S_a =0$ and 
\barray
\nonumber
S_t &=& \frac{\sinh (4\beta \alpha )}{\sinh(4\alpha (i\pi - \beta))} ~S_0 
\\
S_r &=& i \frac{\sin ( 4\pi \alpha )}{\sinh(4\alpha (i\pi - \beta))} ~ S_0
\label{Su1}
\\
\nonumber
S_0 &=&  \frac{\sinh( 2 \alpha (i\pi - \beta ))}{\cosh (2 \alpha \beta )
\sinh(2\pi i \alpha ) } ~ \sigma 
\earray 
Since $\gamma'$ goes to $\infty$, the expression
for $\sigma$ leads to an integral.  Using in addition  the integral:
\beq
\label{integ}
\log \cos (  2\alpha a  ) = - \int_0^\infty \frac{dx}{x} 
\frac{\cosh(ax) -1 }{\sinh( \pi x / 4\alpha )} , 
\eeq
which is valid for $|a| < \pi /4\alpha $,
 one can  represent the additional  trigonometric
factors in eq. (\ref{Su1}) with $c=\pi / 4\alpha$, and one finds: 
\beq
\label{S0int}
-i \log S_0 (\beta ) = \int_0^\infty \frac{dx}{x} 
\frac{ \sin(\beta x) \sinh(( 1/4\alpha -1 ) \pi x/2 )}
{\cosh(\pi x /2) \sinh( \pi x/8 \alpha )}
\eeq
Using eq. (\ref{sGparam}), this 
 agrees with the known sine-Gordon S-matrix\cite{ZZ},  with precisely
the right dependence on the sine-Gordon coupling $b$. 
($S_0$  can also  be expressed as an infinite product of
$\Gamma$ functions; see appendix B.) 

Consider next the limit $\krg = k=0$, with $Q_x \equiv Q$.    In this limit,
\beq
\label{new1}
\alpha = \sqrt{Q}  \equiv \frac{h}{4} 
\eeq
and the period $\lamsG$ remains finite: $\lamsG = 2\pi / h$. 
This theory should correspond to the cyclic sine-Gordon model
of \cite{LRS}.   In order to compare this limit of $S_{\EsG}$ with
the S-matrix in \cite{LRS}, it is necessary to go to the 
$A_{1,2}$ basis of particles with S-matrix $\Sigma$.  
When $k=0$,  the non-zero amplitudes are $\sigma, \sigma_r , \sigma_a$.
As explained in section IV, in order to compare with the usual
cyclic sine-Gordon model, which is defined when $g_x = g_y$, one
must make the $\CP$ transformation which exchanges $g_y$ and $g_z$.  
Making this  $\CP$-transformation to $\Sh$ defined by  eq. (\ref{flip}),
one finds in this limit:
\barray
\nonumber
\Sh_t &=& \frac{ \sin (h \beta/2 )}{\sin(h(i\pi - \beta)/2)} ~ \Sh_0 
\\
\label{cyclelim}
\Sh_r &=& i \frac{\sinh(\pi h/2 )}{\sin(h(i\pi - \beta)/2)} ~ \Sh_0 
\\
\nonumber
\Sh_0 &=& -i \frac{\sin(h (i\pi - \beta)/2 )}{\sinh(\pi h/2 )} 
~ \sigma_{k=0} 
\earray
In this case,  since $\gamma \to \infty$ and $\gamma'$ remains finite, 
 the expression eq. (\ref{3.7}) becomes an infinite
sum rather than an integral:
\beq
\label{Sh0}
-i \log \Sh_0 (\beta) =  \pi + h  \beta/2  + 
\sum_{n=1}^\infty \frac{2}{n} ~ 
\frac{\sin( n \beta h )}{1+\exp( n \pi h)}  
\eeq
The above S-matrix agrees  with the one in \cite{LRS}.

\section{S-matrix for $\EcsG$}

As explained in section IV,  the $\EsG$ and $\EcsG$ models are
essentially different descriptions of the same theory at complimentary
elliptic
moduli.   The discussion therefore closely parallels that of the 
last section, and so we provide less details.  

\subsection{S-matrix}

As we now argue, the  S-matrix for this theory is more   naturally
described by  $\Sh$.  
As for $\EsG$, matching the period $\gamma'$ of the S-matrix with 
the period of the RG $\lamcsG$, $\gamma' = 2\lamcsG$,  leads to:
\beq
\label{3.13}
\frac{\Kb (k)}{\alpha} = \frac{  i  \Kb'  (\krg )}{\sqrt{Q_x}}  
\eeq
Repeating the arguments that led to eq. (\ref{3.12}) one obtains:
\beq
\label{3.14}
\krg' = \frac{2\sqrt{k}}{1+k}  ~~~~~ \sqrt{-Q_x} = (1+k) \alpha 
~~~~~(\EcsG)
\eeq
Note that since $Q_x<0$, $\alpha$ is still real.   
In summary the S-matrix is
\beq
\label{summar2}
S_{\EcsG} (\beta) = \Sh (\beta; k,\alpha) ,
\eeq
where $\Sh (\beta; k, \alpha)$ is defined in eqs. (\ref{Shat2}),
and the parameters $k,\alpha$ given in terms of $Q_x , \krg$ in
eq. (\ref{3.14}).

\subsection{Resonance poles}

The poles of $\sn ( 2\pi i \alpha - 2 \alpha \beta )$ lead
to poles at $\beta = i \pi ( 1 - (2m+1)\gamma/\pi ) + n \gamma' / 2$ 
in the amplitudes $\Sh_{0,a}$.  As for $\EsG$, requiring that
there are no complex poles on the physical strip leads to 
the constraint $\gamma > \pi$. 
Given that the above constraint is satisfied, there remains
resonance poles  at:
\beq
\label{cpoles}
\beta_n = \mu_n - i \eta_n , ~~~~~ 
\mu_n = n \gamma' / 2, ~~~ \eta_n =  \gamma - \pi  
\eeq
Again, for the massive case $n$ must be positive and 
 these resonances only exist for $\gamma < 2 \pi$. 

Noting that $\mu_{n+1} - \mu_n = \gamma' / 2 = \lamcsG$
the Russian doll scaling of the resonances in the UV is:
\beq
\label{dollsc}
M_{n+2} \approx e^{\lamcsG} M_n ,  ~~~~~{\rm for ~ n ~ large} 
\eeq
This is still consistent with the UV signature of
a cyclic RG described in section II, eq. (\ref{dolls}),
the difference being that   {\it two} states are reshuffled
in each RG cycle $\lamcsG$.  

\subsection{$U(1)$ invariant limits}

Again the $U(1)$ invariant limits $\krg = 0,1$ serve as  non-trivial
checks.   
First consider $\krg =1$.  
  In this limit,  $k=0$,  and we parameterize
$\sqrt{Q_x} \equiv \sqrt{Q} \equiv  i h/4$ as in eq. (\ref{2.14}).
  One finds in
this  limit
\beq
\label{cparam}
\alpha = \sqrt{-Q} = h/4 
\eeq
and 
 the period of the RG remains finite, $\lamcsG = 2\pi / h$,
and agrees with the RG calculation in section II.  

The parameter $\gamma$ on the other hand becomes infinite
and the resonances disappear from the spectrum since $\eta > 1$
and they are not on the  strip $-\pi < Im(\beta ) <0$.    
Note also that since $\Kb' (0) = \infty$,
the constraint eq. (\ref{const}) is not required
and $0<h<\infty$.

One finds for the S-matrix the same result as in eq. 
(\ref{cyclelim}), and again this
 S-matrix agrees  with the cyclic sine-Gordon 
one in \cite{LRS}.  

Finally we consider the $\krg =0$ limit.  Here   
the period of the RG goes to infinity  and the theory should
be equivalent to the sine-Gordon model with coupling 
$b$ and $Q \equiv -Q_x$ related as in eq. (\ref{bQ}).  From 
eq. (\ref{3.14}), this limit corresponds to $k=1$ which
leads to 
$\alpha = \sqrt{Q}/2$.  In order to make contact with  the usual sine-Gordon
S-matrix in this limit, we first make the transformation
to $\Sigma = \CP (\Sh)$, then make a change of particle basis 
so that the S-matrix is given by $S$ in equations
(\ref{S.4},\ref{Sexplicit}).   Using now
$\ell = 0, \alphap = 2 i \alpha$, one finds the result in
eq. (\ref{Su1}), with again 
correct dependence on the sine-Gordon coupling $b$.

\section{Scalar theories:  Elliptic sinh-Gordon} 

\def\K{\Kb}

The sinh-Gordon model is defined by the action
\beq
\label{sh.1}
S_{\rm shG} = \inv{4\pi} \int d^2 x  ~\(  \inv{2} (\d \phi)^2 + 
\Lambda \cosh b \phi  \) 
\eeq
As explained in \cite{BLflow}, this 
 model is realized in the field theory of section II when
$Q_x = Q_y \equiv Q$ with $\sqrt{Q} < -1/2$.   The relation
between $Q$ and $b$ is known\cite{BLflow}:   
\beq
\label{shGb}
2 \sqrt{Q} = - (2+ b^2)/b^2 ,
\eeq
 which follows from eq. 
(\ref{bQ}) with $b \to i b$.  

The spectrum of the model consists of a single massive scalar  particle with 
S-matrix\cite{sinhS}:
\beq
\label{sh.2}
S_{\rm shG} = \frac{ \tanh(\beta -i \pi a)/2}{\tanh (\beta + i \pi a)/2 }
\eeq
where 
\beq
\label{sh.3}
a = \frac{b^2}{2+ b^2} 
\eeq

Mussardo and Penati have considered the simplest possible 
scalar S-matrix built out of elliptic functions\cite{Mussardo}:
\beq
\label{sh.4}
S(\beta ;  k, a) = \frac{ \sn (2 \K i \beta /\pi ; k ) + \sn ( 2 \K a ;k) }
{ \sn (2 \K i \beta /\pi ; k ) -  \sn ( 2 \K a ;k  )}
\eeq
where $\K = \K (k)$. 
In the limit where the elliptic modulus $k \to 0$, one recovers
the sinh-Gordon S-matrix eq. (\ref{sh.2}). 

We propose that the above elliptic S-matrix describes the 
elliptic sinh-Gordon regime of the field theory in section II.  
As before, we  relate the S-matrix parameters $k, a$ to the 
lagrangian parameters $Q_x , \krg$ by matching the periodicities.  
The above S-matrix has the periodicity:
\beq
\label{sh.5}
S(\beta - 2 \lambda ; k,a) = S(\beta; k, a) , ~~~~~~
\lambda =  \frac{\pi}{2} \frac{\K' (k) }{\K (k)} 
\eeq
Identifying $\lambda$ as the period of the RG, which from section IV
equals $2\lambda_1$ since $Q_x$ is positive,   one obtains:
\beq
\label{sh.6}
 \frac{\K (\krg ) }{ \sqrt{Q_x}} = -\frac{\pi}{2}    \frac{\K' (k)}{\K (k)} 
\eeq

Since the equation (\ref{sh.6}) does not depend on $a$,  we fix
$a$ using eqns. (\ref{shGb},\ref{sh.3}):
\beq
\label{sh.7}
a = - \inv{2 \sqrt{Q_x}} 
\eeq
The above identification guarantees that the S-matrix has
the correct limit as $k\to 0$.  
Given the above identification of $a$,  then the value of $k$ is 
determined by eq. (\ref{sh.6}):
\beq
\label{sh.8}
\Kb (\krg ) = \frac{\pi}{4a} \frac{\K' (k)}{\K (k)}
\eeq
For $0<\krg <1$, $a>0$,  there is always a solution of the
above equation with $ 0 < k <1$. 
This completes the identification 
of $k,a$ in terms of $Q_x , \krg$. 
Near $\krg = 1$ one has:
\beq
\label{nearo}
k \approx 4 (\krg' /4)^{2a} 
\eeq
so that $k$ approaches $0$ as $\krg \to 1$.    

The above S-matrix has no complex poles on the physical strip 
as long as $0<a<1$.   In terms of the lagrangian parameter this
reads
\beq
\label{const3}
\sqrt{Q_x} < -1/2
\eeq
In the usual sinh-Gordon  limit $k\to 0$, $Q_x \to Q$, the above
constraint correctly goes over to the sinh-Gordon regime
(see section II),  which provides a check on the S-matrix
and eq. (\ref{sh.7}).

The remaining resonance poles are at:
\beq
\label{resshG}
\beta_n = \mu_n - i \eta_n , ~~~~~\mu_n = 2n\lambda, ~~~\eta_n = \pi a
\eeq
Again these resonance poles satisfy the expected UV  Russian doll scaling
eq. (\ref{dolls2}).

\section{Massless elliptic sine-Gordon model ($\mEsG$)}

The only region not covered in the $U(1)$ invariant limit 
by previous cases is $-1/2 < \sqrt{Q} < 0$.   The $U(1)$ invariant
model has an infrared fixed point and is thus massless.
When the model is fully isotropic, $\sqrt{Q} = 0$,  it corresponds
to the large distance limit of the $O(3)$ sigma model at
$\theta = \pi$\cite{ZZmassless}.  In this section we propose 
an S-matrix when $\krg \neq 1$.  The discussion closely parallel's
the $\EsG$ case of section V.

An S-matrix description of massless theories  was
given by Zamolodchikov and Zamolodchikov\cite{ZZmassless}.  
An essential ingredient of their formulation are  
S-matrices for only left-movers or only right-movers, 
denoted $S_{LL}$ and $S_{RR}$.  These S-matrices are formal
S-matrices for a scale-invariant theory.
  When left-right scattering $S_{RL}$ is non-trivial, the
scale invariance is broken.

Since $\sqrt{Q_x}$ is real, the period of the RG is $\lambda = 2\lambda_1$. 
Matching the periodicity $\gamma'$ of the S-matrix while requiring
$\alpha >0$ gives:
\beq
\label{8.1}
\frac{\Kb(k)}{\alpha} = - \frac{ \Kb (\krg )}{\sqrt{Q_x}}  ~~~~~(\mEsG) 
\eeq
The RHS is positive since here $\sqrt{Q_x}$ is negative.  
As for $\EsG$, we argue that the above relation requires:
\beq
\label{8.2}
\sqrt{Q_x} = -(1+k) \alpha , ~~~~~~~~~\krg  = \frac{2\sqrt{k}}{1+k} 
\eeq

  Let us parameterize the massless
energy momentum for left and right movers as in eq. 
(\ref{II.7}).     Requiring the two-particle  S-matrices to correspond
to the $O(3)$ sigma model at $\theta = \pi$ in the $su(2)$ invariant
limit $\krg=1, Q=0$  leads to the obvious proposal:
\barray
S_{RR} (\beta) &=& S_{\EsG} (\beta ) , ~~~~~ \beta = \beta_{R1} - \beta_{R2}
\nonumber 
\\
S_{LL} (\beta) &=& S_{\EsG}  (\beta ) , ~~~~~ \beta = \beta_{L1} - \beta_{L2}
\label{masslessS}
\\
S_{RL} (\beta) &=& S_{\EsG}  (\beta ) , ~~~~~ \beta = \beta_{R1} - \beta_{L2}
\nonumber
\earray
where all the S-matrices $S_{\EsG}$ on the right hand side 
are the same as in eq. (\ref{Sexplicit}), where now
$\alpha, k$ are determined by eq. (\ref{8.2}).  As for other cases,
one can easily check that this has the correct limit as $\krg =1$.

The S-matrix has the periodicity eq. (\ref{II.9}). 
The abscence of complex poles on the physical strip requires
$\gamma > \pi$.  Using eq. (\ref{8.2}) this gives
$\sqrt{Q_x} > - (1+k) \Kb ' (k) / 2\pi $.   Since 
$\Kb' > \pi /2 $,  near $\krg =1$, this gives 
$\sqrt{Q} > -1/2$, which is the expected  range.  

The analysis of resonance poles closely parallels the discussion
in section VI.   The poles eq. (\ref{respoles}) are still present,
but now as explained in section II, $\mu_n$ can be negative.  
This leads to a spectrum of resonances:
\beq
\label{mlessdolls}
M_n = m ~ e^{n\lambda} ~ \cos ((\gamma - \pi )/2) ~~~~~~
n \in \Zmath
\eeq
where the above equation is exact since the theory is massless. 
The above resonances accumulate at zero as  $n \to -\infty$
and are infinitely heavy  as $n\to +\infty$,  which are  both
the UV and IR Russian doll signatures.   This theory is thus
consistent with a cyclic RG on all scales.

\section{Conclusions}

In summary,  based on the limit-cycle behavior of the RG for
maximally anisotropic $su(2)$ current interactions, we have
proposed that they underly the previously known 
exact S-matrices built out of elliptic functions.   
The S-matrix and lagrangian parameters were related by
matching the period of the RG with the UV signature
of a cyclic RG in the S-matrix, i.e. the periodicity
in rapidity.   Numerous checks were performed, in particular
we showed the models have an infinite spectrum of resonances
consistent with the cylic RG 
 and  have shown that the
models have the expected  $U(1)$ invariant limits.

Since in this paper we have proposed a field theory for the 
elliptic S-matrices for the first time,  there are many open
avenues for further investigation, and we finish this paper by
listing a few of them. 

The sine-Gordon theory,  which appears in the trigonometric limit
of our model,  is classically integrable, 
and in fact semi-classical methods using this integrability were
used early on\cite{Dashen}  to study the spectrum of the model and these
results eventually provided some checks on the S-matrix\cite{ZZ}.    
To our knowledge the classical (and quantum) integrability
of our field theory has not been studied.  Clearly we have
assumed it was integrable in proposing exact S-matrices. 
A good starting point is the bosonized action (\ref{Sbosonic}).

The sine-Gordon theory which arises in the trigonometric limit
is known to have a quantum affine $ \hat{sl(2)_q}$ symmetry
and the conserved charges can be constructed explicitly in the
quantum field theory\cite{nonlocal}.   It would be interesting to
extend this field theory construction to the present models
since this should lead to an elliptic deformation of the affine
$\hat{sl(2)}$ algebra which can be compared with the algebra in
\cite{MiwaJimboElliptic}.

Our model may have some applications to solid-state physics.
In the fermionic representation of the $su(2)$ currents, the model
is essentially a Luttinger liquid for fermions with spin
and  2 additional kinds of 
density-density perturbations since  the usual Luttinger liquid
corresponds to $g_x = g_y = 0$.  

Theories with an infinite number of resonances are reminiscent of
string theories\cite{strings}.  In string theory the resonances
are exactly stable,  whereas in our model they generally have
a finite lifetime.  Another S-matrix was studied in \cite{LRS} which,
though related, is essentially different from the S-matrices
here,  and is characterized by an infinite number of exactly
stable resonances but with no periodicity in rapidity.  A field
theory interpretation of this S-matrix seems unlikely since
it suffers from a lack of real analyticity.   How this S-matrix
is related to the physical S-matrices in this paper is described
in appendix B.

\section*{Acknowledgments}

We would like to thank G. Mussardo and S. Pakuliak  for discussions. 
This work has been supported by the Spanish grants 
SAB2001-0011 (AL), 
BFM2000-1320-C02-01 (GS), and by the NSF of the USA. 
We also thank the EC Commission for financial support
via the FP5 Grant HPRN-CT-2002-00325.

\section{Appendix A: Higher orders}

In \cite{GLM} an all orders beta function was proposed for 
general anisotropic current interactions.  In addition
to the coefficients $C$ defined in eq. (\ref{2.3}) the
beta function depends on the following.  Expressing 
 the perturbing operators as $\CO^A = d^A_{ab} J^a \Jbar^b$,
 define the purely chiral  operators
\beq
\label{A.1}
T^A (z) \equiv d^A_{ab} J^a (z) J^b (z) 
\eeq
In the conformal field theory one has the closed operator product
expansion
\beq
\label{A.2}
T^A (z) \CO^B (0) \sim \inv{z^2} \( 2 \kh  D^{AB}_C + \tilde{C}^{AB}_C \) 
\CO^C (0) 
\eeq
The formula in \cite{GLM} for the beta function is then expressed
in terms of the coefficients $C, \tilde{C} , D$.   

For our model one easily finds the non-zero values:
\beq
\label{A.3}
D^{xx}_x = D^{yy}_y = D^{zz}_z = 2 
\eeq
\beq
\label{A.4}
\tilde{C}^{xy}_y = \tilde{C}^{xz}_z = 
\tilde{C}^{yx}_x =\tilde{C}^{yz}_z =  \tilde{C}^{zx}_x = \tilde{C}^{zy}_y 
= 4
\eeq  From the formula given in \cite{GLM} one finds
\beq
\label{A.5}
\beta_z  =  \frac{4(   g_x g_y(1+\kh^2 g_z^2 ) - \kh g_z (g_x^2 + g_y^2 ) )} 
{(1-\kh^2 g_x^2 ) (1-\kh^2 g_y^2 )}
\eeq
where here $\kh$ is the level of the current algebra, which for our
model equals 1. 
The other two beta functions $\beta_x , \beta_y$ follow from the
above expression by permutation of the $x,y,z$ indices.

The above beta function has a strong-weak coupling duality.   For the
$U(1)$ invariant case $g_x^2  = g_y^2 $, this duality was exploited in
\cite{BLflow} in order to extend the flows to all scales.  Since
the level can be easily scaled out of the equations,  let us set
$\kh =1$.   
For each coupling,  define the dual coupling as $g^* = 1/g$,
and the beta function for the dual couplings:
\beq
\label{dual1}
\beta^* (g^*) =  \beta (g)  \frac{dg^*}{dg} 
\eeq
Then one can verify that the beta function satisfies the duality
relation:
\beq
\label{dual2}
\beta^* (g^*) = - \beta(g \to g^*)
\eeq
An important consequence of the duality  is that flows between $g$ equal to
$0$ and $1$ 
can be mapped onto flows between $g$ equal to $1$ and $\infty$.
We can now  use
this duality to argue that the flows based on the above
all-orders beta functions continue to be cyclic as follows.  At the
intial RG time, 
suppose that the couplings are near zero, and that running forward in 
RG time they reach $g=1$ whereas running backwards they reach $g=-1$.  
Though the points $g=\pm 1$ are poles in the beta functions, it was
shown in the $U(1)$ invariant case that because of the RG invariants
the flows approach the poles along tangent directions determined
by $Q$ and flow smoothly through the pole.   In other words,  the
poles are not true singular points:  a local blow up resolves
the flows.   Beyond $g=1$, the flow between $g=1$ and $\infty$ is
dual to the flow between $0$ and $1$.    Thus it takes the same time
to flow between $1$ and $\infty$ as it does between $0$ and $1$.  
At $g= \infty$ the flow actually continues smoothly at $g=-\infty$; 
this jump is exactly dual to a smooth flow through a $g=0$ since
$1/0^{\pm} = \pm \infty$.    The flow then continues to $g=-1$ and
a new cycle begins.  
The period of the RG is then twice
the time it takes to flow between $-1$ and $1$.         
We will use this below to numerically determine the RG period.

What complicates the situation in the fully anisotropic case
is that unlike the $U(1)$ invariant one  we have been unable to
find simple  expressions for the higher order RG invariants
$Q$, nor have we been able to integrate the all orders beta function
and compute the period $\lambda$ analytically. 
However we can at least 
show that the existence of the
RG invariants is 
not spoiled at 2-loops.
Additionally, the 
 two-loop analysis shows that the corrections to the $Q$'s are  not
 simple power series.  
   Let $\beta (g) = \sum_{i=1}^\infty  \beta^{(i)} $ where
$\beta^{(i)}$ is the i-th loop contribution the beta function, of
order $g^{i+1}$.  Let us also write  for $Q= Q_{x,y,z}$:
\beq
\label{higher2}
Q  = Q^{(1)} + Q^{(2)} + .\, . \, .\, . 
\eeq
where 
$Q^{(1)}$ is the one-loop expression given in eq. (\ref{2.6}),
and $Q^{(2)}$ is the two loop correction.
Assuming $Q^{(2)}$ is of order $g^3$, 
the RG invariance of $Q$ to two loops requires
\beq
\label{higher3}
\sum_{i=x,y,z} \beta_i^{(1)} \d_{g_i} Q^{(2)} + 
\beta_i^{(2)} \d_{g_i} Q^{(1)}
=0      
\eeq
Keeping just the two-loop contributions to eq. (\ref{A.5}) one finds 
\beq
\label{higher4}
\beta_z = 4g_x g_y - 4g_z (g_x^2 + g_y^2)+ . \, .\, .\, . 
\eeq
and $\beta_{x,y}$ again  given by the obvious permutations.  Using this in
eq. (\ref{higher3}) one finds the following two loop correction to
$Q_x$:
\beq
\label{higher5}
Q_x =  (g_z^2 - g_y^2 ) \[  1+ 2 \int^{g_x}   \frac{u^2 du}
{\sqrt{(u^2 + g_y^2 - g_x^2 )(u^2 + g_z^2 - g_x^2  )} } \]  + ....
\eeq
The above  is an elliptic integral of the third kind.
This suggests that the RG flows can be uniformized using
elliptic functions, reminiscent of Seiberg-Witten theory\cite{seiberg}.

We now  can  give numerical evidence for one of the main hypotheses
of this paper, i.e. that the period of the RG is twice
the 1-loop result, eq. (\ref{lams}).   If the couplings
are initially very small, then we expect that we can approximate
the $Q$'s by their one loop expressions $Q^{(1)}$.    The results in the 
case $Q_x < 0$ for several values of initial couplings 
 are shown in 
Table 1. 
The analytic expression   $\lambda' = 2\lambda'_1$, is given by
the formula (\ref{2.12}),  where $Q_x, Q_y $ are approximated
by the 1-loop result $Q_{x,y}^{(1)}$.
 The last column, $\lambda_{\rm num}$,  
denotes the period of the exact RG evolution using the
beta functions eq. (\ref{A.5}),  
which we have computed
numerically as the time to go from a $g_z=-1$ to $g_z = 1$. 
In the $U(1)$ case, where $g_x = g_y=0.1$, there is a small discrepancy
between   $2\lambda'_1$ and  $\lambda_{\rm num} $ which
is an indication that we have used the 1-loop expressions
$Q^{(1)}$ in the expression for $\lambda_1$. 
This  presumably explains also explains the small discrepancy when
$\krg \neq 1$.

\bigskip

\begin{center}
\begin{tabular}{|c|c|c|c|c|c|c|c|}
\hline
$g_x$ & $g_y$ & $g_z$ & $\krg$ & $Q_x^{(1)}$ &$Q_y^{(1)}$ & 
$2\lambda'_1$ & $\lambda_{\rm num} $ \\
\hline 
0.1 & 0.1 & 0 & 1 & -0.01 & -0.01 & 15.708 & 15.629 \\
0.1 & 0.2 & 0 & 0.5 & -0.04 & -0.01 &  10.783 &  10.660 \\
0.1 & 0.3 & 0&  0.333 & -0.09 & -0.01 &  8.429 &  8.2590 \\
\hline
\end{tabular}

\vspace{0.5 cm}
Table 1.- Numerical comparison between the analytic expression  $2\lambda'_1$
and the numerically determined period   $\lambda_{\rm num}$. 

\end{center}

\section{Appendix B:  Infinite products of $\Gamma$-functions
and a  stringy S-matrix}

The other S-matrix  considered in \cite{LRS} was an analytic extension 
of the usual sine-Gordon one to the complex values of
the sine-Gordon coupling $b$:
\beq
\label{B.1}
2 \sqrt{Q} = \frac{2-b^2}{b^2} = \frac{ih}{2} 
\eeq
As pointed out in \cite{LRS},  the resulting S-matrix
is not real analytic:  $S^\dagger (\beta ) \neq S(-\beta)$, 
and for this reason more than likely does not have a field theory
description.   Regardless,  the S-matrix has some interesting
properties.  It is characterized by a Russian doll spectrum 
of resonances in the UV but with no periodicity in rapidity. 
The resonances
are exactly stable,  and closing the bootstrap led to a string-like 
spectrum\cite{LRS}.

This stringy S-matrix has not played any role in this paper,
nevertheless it is closely related to the $\EsG$ S-matrix with
$\alpha$ purely imaginary,  
as we now explain.  
Consider the $\EsG$ S-matrix in the limit $k\to 1$,  which
leads to eq. (\ref{Su1}).   
Let $\alpha$ be defined by eqns. (\ref{sGparam},\ref{B.1}):
\beq
\label{B.2}
\alpha = \frac{ih}{8}
\eeq
Then the S-matrix in eq. (\ref{Su1}) has the overall structure
of ratios of the cyclic sine-Gordon model  
in eq. (\ref{cyclelim}) with $\Sh \to S$. 
      When $k\to 1$,  $\gamma'$ becomes
infinite and $\gamma$ remains finite,  thus the expression 
for $\sigma$ eq. (\ref{3.7}) becomes an integral.  However
the integral does not converge for $\alpha$ purely imaginary. 

\def\Ga{\Gamma}
\def\rap{\beta}

In \cite{LRS}  a  different overall scalar factor was proposed 
which is just the analytic extension of the usual sine-Gordon one. 
Define:
\beq
\label{B.3}
S_0^\Gamma  = \frac{ \Gamma (4\alpha ) \Gamma (1-z)}{\Gamma(4\alpha -z ) }
\prod_{n=1}^\infty  
\frac{ F_n (\beta  ) F_n (i\pi - \beta )}{F_n (0) F_n (i\pi ) }
\eeq
where
\beq
\label{B.4}
F_n (\rap ) = \frac
{ \Ga \( 8n\alpha   - z \) \Ga \( 1 + 8n\alpha  -z \) }
{\Ga \( 4(2n+1)\alpha   - z \) \Ga \( 1 + 4(2n-1)\alpha -z\) }. 
\eeq
and we have defined:
\beq
\label{B.5}
z = - \frac{4i\beta \alpha}{\pi}
\eeq
When $\alpha$ is real and related to the sine-Gordon coupling $b$
as in eq. (\ref{sGparam}), the above is the well-known expression 
for $S_0$ as an infinite product of $\Gamma$-functions.  
(We will show it is equivalent to  
eq. (\ref{S0int}) when $\alpha$ is real below.)

Consider now the above infinite product when $\alpha$ is pure imaginary.
The product is still convergent.     It can be given
an integral representation using:
\beq
\label{B.6}
\int_\CC  \frac{dx}{2\pi i x}  \log (-\pi x)   
\frac{ e^{-a \pi x}}{1- \exp ( -\pi x /4\alpha )} 
= \log \Gamma (4\alpha a ) + (4\alpha a   - 1/2 ) (\gamma - \log 4\alpha
 ) -  \log (2\pi)/2 
\eeq
where $\gamma$ is Euler's constant, and the contour $\CC$ is shown
in figure 2.   The above integral is valid for $\alpha$
 real or purely imaginary
as long as  the real part of $a$ positive.
Nearly all the $\Gamma$ functions in $\log S^\Gamma_0$ can be represented 
with the real part of
$a$ a positive integer.  The sum over $n$ converges since
$\sum_{n>0}  \exp (-2n\pi x)$ converges. The result is:
\beq
\label{S0gamma}
S_0^\Gamma = \frac
{\Gamma ( 1+ 4i \beta \alpha/\pi ) \Gamma (4\alpha - 4i\beta \alpha/\pi )}
{\Gamma ( 1- 4i \beta \alpha/\pi ) \Gamma (4\alpha + 4i\beta \alpha/\pi )}
~ I 
\eeq
where
\beq
\label{logs}
\log I  = \int_\CC \frac{dx}{2\pi x} \log (-\pi x) ~ 
\frac
{\sin (\beta x) \sinh (\pi x (1-1/4\alpha )/2 ) }
{\cosh (\pi x/2 ) \sinh (\pi x / 4\alpha ) } ~  e^{-\pi x}
\eeq

In eq. (\ref{S0gamma}), we have factored out the $\Gamma$ functions 
that cannot be represented by an integral when $\alpha$ is imaginary. 
The above integral is convergent which proves the infinite product
is convergent.

\begin{figure}[h]
\begin{center}
\includegraphics[height=6 cm, angle=0]{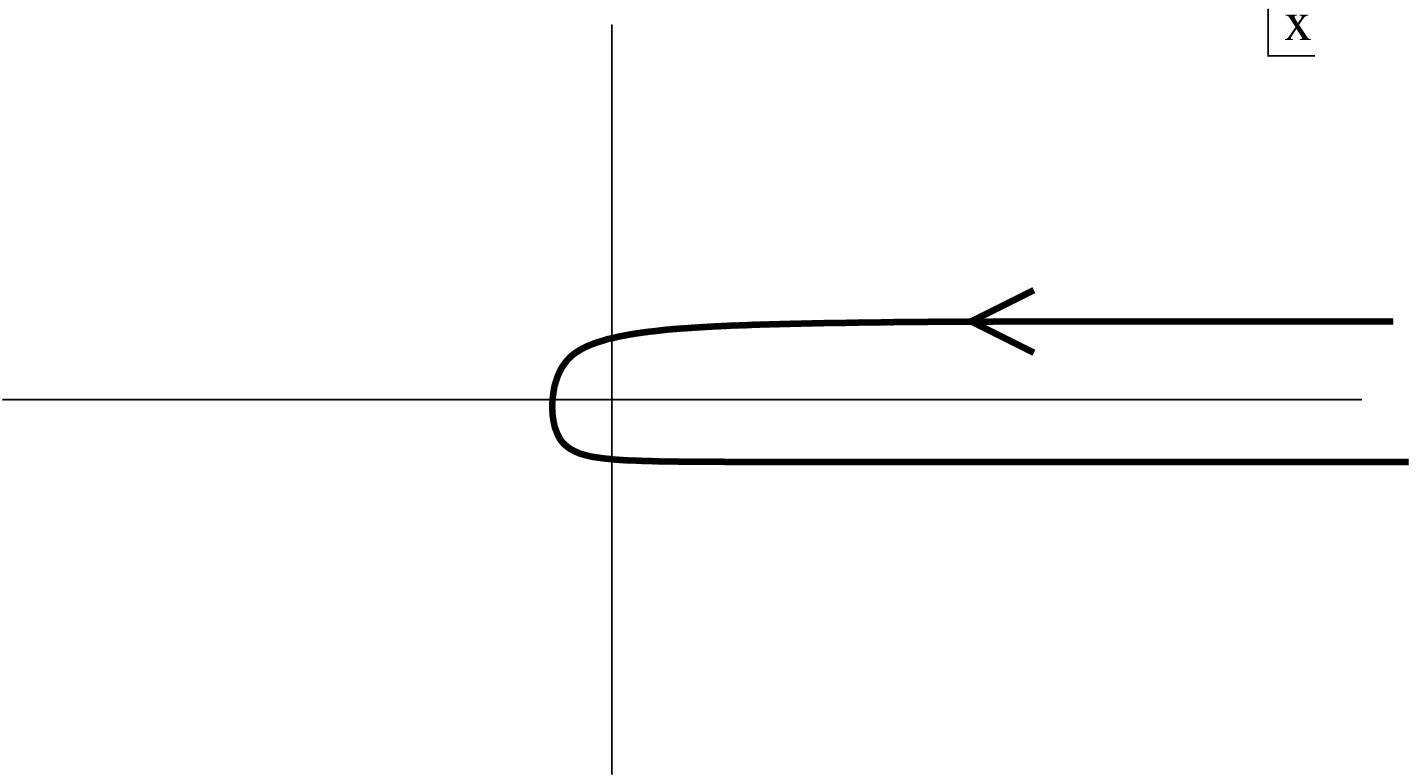}
\end{center}
\caption{Integration contour $\CC$ for eq. (\ref{B.6}).  } 
\label{fig2}
\end{figure}

When $\alpha$ is real, the additional $\Gamma$ functions 
in eq. (\ref{S0int}) can also be represented by a contour integral.
Furthermore, since in this case  there are no poles on the real $x$ axis, 
  the contour integral can be replaced
by an ordinary integral:
\beq
\label{B.8}
\int_\CC  
  \frac{dx}{2\pi i x}
\log (-\pi x )    ~ \longrightarrow  \int_0^\infty  \frac{dx}{x} 
\eeq
and one recovers eq. (\ref{S0int}).  For $\alpha$ imaginary
however,  because of the $1/ \sinh (\pi x / 8\alpha )$ in
the integrand one cannot safely make the replacement eq. (\ref{B.8}). 

In summary, when $\alpha$ is imaginary, one cannot obtain
the convergent expression eq. (\ref{S0gamma}) from the 
$k\to 1$ limit of $\sigma$ in eq. (\ref{3.7}).  Rather, 
one has to perform the integrals in eq. (\ref{S0int}) 
for $\alpha$ real obtaining the infinite product of
$\Gamma$ functions, and then analtyically continue to
imaginary $\alpha$.

\vfill\eject


\begin{thebibliography}{99}




\bibitem{ZZ} A. B. Zamolodchikov and  Al. B. Zamolodchikov,
``Factorized S-matrices in Two Dimensions as the Exact 
Solutions of Certain Relativistic Quantum Field Theory
Models'', Annals of Phys. {\bf 120}, 253 (1979).   



\bibitem{Zelip} A. B. Zamolodchikov, 
``$\Zmath_4$-symmetric factorized S-matrix in two space-time dimensions'',
Commun. Math. Phys. {\bf 69} (1979) 165. 



\bibitem{Mussardo}  G. Mussardo and S. Penati,
``A quantum field theory with infinite resonance states'',
Nucl. Phys. {\bf B567} (2000) 454, hep-th/9907039. 


\bibitem{Baxter}  R. J. Baxter,  Ann. Phys. {\bf 70} (1972) 193; 
 ``Exactly Solved Models in Statistical Mechanics'', Academic Press, 
London (1982).



\bibitem{Luther}  A. Luther, Phys. Rev. {\bf B14} (1976) 2153. 


\bibitem{BLflow}  D. Bernard and  A. LeClair, 
``Strong-weak coupling duality in anisotropic current interactions'', 
Phys.Lett. B512 (2001) 78; hep-th/0103096. 



 \bibitem{nuclear} P. F. Bedaque, H.-W. Hammer, and U. van Kolck,
``Renormalization of the Three-Body System with Short-Range
Interactions'', Phys. Rev. Lett. {\bf 82} (1999) 463, nucl-th/9809025.


\bibitem{GW} S. D. Glazek and K. G. Wilson, ``Limit cycles in
quantum theories'',  
Phys.Rev.Lett. {\bf 89}  (2002) 230401, 
hep-th/0203088;
``Universality, marginal operators, and limit cycles,''
cond-mat/0303297.



\bibitem{LRS} A. LeClair, J.M. Roman and G. Sierra,
``Russian Doll Renormalization Group 
and Kosterlitz-Thouless Flows'', 
Nucl.Phys. {\bf B675}  (2003) 584,   hep-th/0301042. 



\bibitem{BCS}  A. LeClair, J.-M. Roman and G. Sierra, 
``Russian Doll Renormalization Group and Superconductivity'', 
Phys. Rev. {\bf B69} (2004) 20505,  cond-mat/0211338. 


\bibitem{QCD}   E.  Braaten and  H.-W. Hammer, 
``An Infrared Renormalization Group Limit Cycle in QCD'',
Phys.Rev.Lett. 91 (2003) 102002, nucl-th/0303038. 


\bibitem{LRSc} A. LeClair, J.M. Roman and G. Sierra,
``Log periodic behavior of finite size effects in field theories
with RG limit cycles'', hep-th/0312141.



\bibitem{KWilson}  K. G. Wilson, ``Renormalization Group and 
Strong Interactions'', Phys. Rev. {\bf D3} (1971) 1818.



\bibitem{KZ}  V. G. Knizhnik and A.B. Zamolodchikov, 
``Current algebras and the Wess-Zumino model'', 
Nucl. Phys. {\bf B247}, 83 (1984).  

\bibitem{Witten} E. Witten,
``Non-abelian bosonization in two dimensions'', 
Commun. Math. Phys. {\bf 92} (1984) 455. 


\bibitem{GoddardOlive} For a review, see P. Goddard and D. Olive,
Int. J. Mod. Phys {\bf A1} (1986) 303;  P. Ginsparg, in 
``Fields, strings and critical phenomena'', Les Houches lectures
1988, Eds. E. Br\'ezin and J. Zinn-Justin, North-Holland 1990. 

\bibitem{tsvelik}  A. O. Gogolin, A. A. Nersesyan and A. M. Tsvelik,
``Bosonization and Strongly Correlated Systems'', 
Cambridge University Press (Cambridge, 1998). 


\bibitem{Boyanovsky} D. Boyanovsky and R. Holman, 
``Critical behaviour and duality in extended sine-Gordon theories'', 
Nucl. Phys. B 358, 619 (1991).

\bibitem{Aldo} G. Delfino,
``Off critical correlations in the Ashkin-Teller model'', 
Phys.Lett. {\bf B450} (1999) 196 (hep-th/9811215); 
``Field Theory of scaling lattice models: the Potts
antiferromagnet'', in  {\it Statistical field theories}, A. Cappelli and G.
Mussardo eds., Kluwer Academic Publishers, 2002 ( hep-th/0110181);
G. Delfino and P. Grinza, 
``Universal ratios along a ling of critical points: the
Ashin-Teller model'', 
Nucl. Phys. {\bf B682} (2004)
521 (hep-th/0309129).  
 



\bibitem{Lecheminant} P. Lecheminant, A.O. Gogolin and A.A. Nersesyan, 
``Criticality in self-dual sine-Gordon models'', 
Nucl.Phys. B 639, 502 (2002);
cond-mat/0203294. 

\bibitem{SierraKim} G. Sierra and E.K. Kim, ``Renormalization group study of
the sliding Luttinger liquids'', J. Phys A: Math. Gen. {\bf 36} (2003) L37.



\bibitem{ZamoRG}  A. B. Zamolodchikov, 
``Renormalization group and perturbation theory about fixed
points in two-dimensional field theory'',
Sov. J. Nucl. Phys. {\bf 46} (1987) 1090.


\bibitem{GR}  I. S. Gradshtein and I. M. Ryzhik, 
``Table of integrals, series, and products'', 
Academic Press 1980, New York; M. Abramowitz and I. A. Stegun, 
``Handbook of Mathematical Functions'', Dover Pubs. Inc. 1972, New York.  
 


\bibitem{GLM} B. Gerganov, A. LeClair and M. Moriconi, 
``On the Beta Function for 
Anisotropic Current Interactions in 2D'', 
Phys. Rev. Lett. {\bf 86} (2001) 4753; hep-th/0011189.  


\bibitem{Ramond} M. Peskin and D. Schroeder, 
``An Introduction to Quantum Field Theory'',
Addison-Wesley, 1995;  P. Ramond, 
``Field Theory, A Modern Primer'', Frontiers in Physics {\bf 51}, 
 Benjamin/Cummings 1981. 




\bibitem{braz}  V. A. Brazhnikov, $\Phi^{(2)}$ Perturbations of WZW Model, 
 Nucl. Phys. {\bf B501} 
(1997) 685, hep-th/9612040. 


\bibitem{mira2}  J. L. Miramontes and C. R. Fern\'andez-Pousa,
``Integrable quantum field theories with unstable particles'',  
Phys. Lett. {\bf B472} (2000) 392, hep-th/9910218.

\bibitem{Fring}   O.A. Castro-Alvaredo, J. Dreissig and  A. Fring,
``Integrable scattering theories with unstable particles'', 
hep-th/0211168.

\bibitem{Fring2}
 O.A. Castro-Alvaredo  and  A. Fring,
``Constructing Infinite Particle Spectra,''
    Phys. Rev. {\bf D64} (2001 085005,    hep-th/0103252;
``Breathers in the elliptic sine-Gordon model'',
     J. Phys. {\bf A36}  (2003)  10233,     hep-th/0303245. 



\bibitem{sinhS}  A. E. Arinshtein, V. A. Fateev and A. B. Zamolodchikov,
Phys. Lett. {\bf 87B} (1979) 389. 


\bibitem{ZZmassless}  A. B. Zamolodchikov and Al. B. Zamolodchikov,
``Massless factorized scattering and sigma models with topological
terms'',  
Nucl.Phys.B379 (1993) 521. 


\bibitem{Dashen}  R. Dashen, B. Hasslacher and A. Neveu, 
Phys. Rev. {\bf D11} (1975) 3424.

\bibitem{nonlocal}  D. Bernard and A. LeClair, ``Quantum Group
Symmetries and Non-Local Conserved Currents in 2D QFT'', 
Comm. Math. Phys. {\bf 142} (1991) 99. 


\bibitem{MiwaJimboElliptic}  
Omar Foda, K. Iohara, M. Jimbo, R. Kedem, T. Miwa, and H. Yan, 
``An elliptic quantum algebra for $\widehat{sl}_2$'', 
Lett.Math.Phys. 32 (1994) 259, 
hep-th/9403094.


\bibitem{strings}  M. B. Green, J. H. Schwarz and E. Witten,
``Superstring Theory'', Cambridge University Press (Cambridge 1987).

\bibitem{seiberg}  N. Seiberg and E. Witten, 
``Monopole Condensation, 
And Confinement In N=2 Supersymmetric Yang-Mills Theory'', 
Nucl.Phys. {\bf B426}  (1994) 19; Erratum-ibid. {\bf B430}  (1994) 485;
hep-th/9407087.



\end{thebibliography}
\end{document}